\documentclass[journal]{new-aiaa}
\usepackage[utf8]{inputenc}
\usepackage{textcomp}
\usepackage{cleveref}
\DeclareUnicodeCharacter{221E}{$_\infty$}
\usepackage{graphicx}
\usepackage{amsmath}
\usepackage{mathtools}
\usepackage{bm}
\usepackage[version=4]{mhchem}
\usepackage{siunitx}
\usepackage{longtable,tabularx}
\usepackage{gensymb}
\usepackage{xcolor}
\usepackage{setspace}
\usepackage{multirow}

\DeclareMathOperator\erf{erf}
\renewcommand{\arraystretch}{0.7} 
\title{Uncertainties and Design of Active Aerodynamic Attitude Control in Very Low Earth Orbit}

\author{Sabrina Livadiotti \footnote{Corresponding author: Ph.D. Researcher, Department of MACE, sabrinalivadiotti@gmail.com}, Nicholas H. Crisp\footnote{Postdoctoral Researcher, Department of MACE}, Peter C. E. Roberts \footnote{Senior Lecturer, Department of MACE.} and Vitor T. A. Oiko\footnote{Postdoctoral Researcher, Department of MACE} }
\affil{The University of Manchester, Oxford Road, Manchester, M13 9PL, United Kingdom}
\author{Simon Christensen\footnote{Systems Engineer}}
\affil{GomSpace A/S, Langagervej 6, 9220 Aalborg East, Denmark}
\author{Rosa Maria Dom\'{i}nguez\footnote{Physicist}}
\affil{Elecnor Deimos Satellite Systems, Calle Francia 9, 13500 Puertollano, Spain}
\author{Georg H. Herdrich\footnote{Head Plasma Wind Tunnels and Electric Propulsion, IRS}}
\affil{Institute of Space Systems (IRS), University of Stuttgart, Pfaffenwaldring 29, 70569 Stuttgart, Germany}
\begin{document}

\maketitle

\begin{abstract}
This paper discusses the design and the performance achievable with active aerodynamic attitude control in very low Earth orbit, i.e. below 450 km in altitude. A novel real-time algorithm is proposed for selecting the angles of deflection of aerodynamic actuators providing the closest match to the control signal computed by a selected control law. The algorithm is based on a panel method for the computation of the aerodynamic coefficients and relies on approximate environmental parameters estimation and worst-case scenario assumptions for the re-emission properties of space materials. Discussion of results is performed by assuming two representative pointing manoeuvres, for which momentum wheels and aerodynamic actuators are used synergistically. A quaternion feedback PID controller implemented in discrete time is assumed to determine the control signal at a sampling frequency of 1 Hz. The outcome of a Monte Carlo analysis, performed for a wide range of orbital conditions, shows that the target attitude is successfully achieved for the vast majority of the cases, thus proving the robustness of the approach in the presence of environmental uncertainties and realistic attitude hardware limitations. 

\end{abstract}

\section*{Nomenclature}

{\renewcommand\arraystretch{1.0}
\noindent\begin{longtable*}{@{}l @{\quad=\quad} l@{}}
$\boldsymbol{a_{A}}$& aerodynamic acceleration vector \\
$A^{p,i}_{B}$  & rotation from the i-th panel to the body reference frame \\
$\boldsymbol{a^{w,q}_{B}}$& orientation of the q-th wheel spin axis in body axes \\
$c_{d}$ & diffuse reflectivity coefficient \\
$\boldsymbol{C_{M}}$ & adimensional aerodynamic momentum coefficient vector \\
$\boldsymbol{C_{M_{d}}}$ & dimensional aerodynamic momentum coefficient vector \\
$c_{p}$ & normal pressure coefficient \\
$c_{s}$ & specular reflectivity coefficient \\
$c_{\tau}$ & shear stress coefficient \\
$E_{Q}$& equation of time \\
$f$& Earth's flattening coefficient \\
$h$ & altitude \\
$\boldsymbol{H_{tot}}$& total angular momentum of the system\\
$\boldsymbol{H_{s}}$& angular momentum of the satellite\\
$\boldsymbol{H_{p}}$& angular momentum of the panels\\
$\boldsymbol{H_{w}}$& angular momentum of the wheels\\
$\boldsymbol{J_{s}}$  & time-varying inertia matrix of the satellite \\
$\boldsymbol{J_{p,i}}$ &  inertia matrix of the i-th panel \\
$\boldsymbol{J_{w,q}}$& spin axis moment of inertia of the q-th wheel \\
$K_{n}$ & Knudsen number\\
$LHA_{\odot}$ & local hour angle\\
$l_{ref}$ & reference length \\
$l_{s}$& panels support length\\
$m$& satellite mass\\
$M_{\odot}$ & mean anomaly of the Sun \\
$m_{b}$& residual dipole moment\\
$M_{j}$ & j-th triangular mesh element\\
$m_{m}$& molecular mass of atmospheric constituents\\
$M_{tot}$ &total number of elements in the mesh\\
$\boldsymbol{\hat{n}}$& outward normal unit vector \\
$p$ & i-th panel \\
$P_{N}$ & angular position of the panel (nominal minimum drag)\\
$P_{R}$ & angular position of the panel (non-nominal)\\
$q_{d}$ & dynamic pressure \\
$R$ & universal constant of gas  \\
$\boldsymbol{r_{\odot}}$ & Sun's position vector \\
$\boldsymbol{r_{CoM}}$ & position vector of satellite CoM with regards to the geometric centre \\
${r_{e}}$ & Earth's equatorial radius \\
$\boldsymbol{r_{I}}$ & satellite inertial position vector \\
$\boldsymbol{r_{j}}$ & distance between the barycenter of the j-th element and the geometric centre of the satellite\\
$\boldsymbol{r_{PO}}$ & distance between the centre of mass and the centre of pressure  \\
$s$ & molecular speed ratio \\
$S_{j}$ & surface of the j-th element \\
$S_{ref}$ & reference surface \\
$t_{0}$ & initial epoch in ephemeris seconds past J2000 \\
$\boldsymbol{T_{a}}$& aerodynamic torques vector \\
$T_{alt}$ & temperature at altitude \\
$\boldsymbol{T_{d}}$& disturbance torques vector \\
$\boldsymbol{T_{e}}$& external torques vector \\
$\boldsymbol{T_{i}}$& incident particles temperature \\
$\boldsymbol{T_{r}}$& reflected particles temperature \\
$t_{s}$ & sampling time \\
$\boldsymbol{T_{w}}$& surface temperature \\
$\boldsymbol{u_{a}}$& aerodynamic control signal\\
$\boldsymbol{u_{w}}$& reaction wheel control signal\\
$\varv_{b}$ & gas bulk velocity\\
$\boldsymbol{\varv_{I}}$ & satellite inertial velocity vector\\
$\boldsymbol{\varv_{rel}}$ & satellite velocity vector relatively to the flow\\
$\boldsymbol{\varv_{rot}}$ & atmospheric corotation velocity vector \\
$\varv_{t}$ & atmospheric particle thermal speed\\
$\boldsymbol{\varv_{w}}$ & atmospheric winds velocity vector \\
$w_{mb}$& main bus width\\
$w_{Q}$& weighting coefficients for the state space\\
$w_{R}$& weighting coefficients for the control input\\
$x$& state space vector\\
$x_{e}$& error state space vector\\
$x_{pi}y_{pi}z_{pi}$ & i-th panel reference frame \\
$x_{ref}$& reference state space vector\\
$X_{B}Y_{B}Z_{B}$ & body reference frame \\
$Y_{B,m}$, $Z_{B,m}$ & Y and Z body coordinates of the m-th mesh element vertex \\
$Y_{CoM}$, $Z_{CoM}$ & Y and Z body coordinates of the CoM \\
$\alpha_{k}$ & angle of attack \\
$\alpha_{n}$ & normal thermal energy accommodation coefficient \\
$\alpha_{T}$ & thermal energy accommodation coefficient \\
$\beta_{k}$ & angle of sideslip \\
$\gamma_{p}$ & weighting coefficient \\
$\delta$ & angle of incidence of the flow (measured from the normal to the surface) \\
$\epsilon$ & obliquity of the ecliptic\\
$\vartheta_{max}$ & saturation angle of deflection of the panels \\
$\vartheta_{p,i}$ & angle of deflection of the i-th panel \\
$\vartheta_{pl}$ & panels minimum angle of deflection (plant) \\
$\vartheta_{s}$ & panels minimum angle of deflection (algorithm) \\
$\lambda_{E_{\odot}}$ & ecliptic longitude of the Sun \\
$\lambda_{M_{\odot}}$ & mean longitude of the Sun \\
$\nu_{M}$ & mean anomaly \\
$\rho$ & thermospheric density \\
$\sigma_{n},\sigma_{t}$ & normal/tangential momentum accommodation coefficient \\
$\boldsymbol{\hat\tau}$& unit tangent vector \\
$\phi$ & roll angle \\
$\phi_{gd}$ & geodetic latitude\\
$\phi_{l}$ & reference longitude\\
$\varphi$ & pitch angle \\
$\psi$ & yaw angle \\
$\boldsymbol{\omega^{I}_{B}}$ & inertially referenced body angular rates \\
$\boldsymbol{\omega_{p,i}}$ & angular velocity vector of the i-th panel relatively to the satellite bus \\
$\boldsymbol{\omega_{w,q}}$ & angular velocity of the q-th wheel about its spin axis \\
$\boldsymbol{\omega_{\oplus}}$ & Earth's angular velocity vector \\
\end{longtable*}}

\section{Introduction}
\lettrine{T}{he} majority of space missions in low Earth orbit (LEO) are characterised by operational altitudes spanning from 600 to 2000 km. The lower altitude range - extending below 600 km - is generally avoided for practical applications due to the numerous challenges that the enhanced disturbance environment poses to the platform design. Satellites travelling in these orbits experience increased magnetic dipole interactions, gravity gradient, aerodynamic torques, and similar effects due to solar radiation pressure present at higher altitude orbits. This is especially true in the very low Earth orbit range (VLEO), i.e. below 450 km, where the order of magnitude of the aerodynamic torques is the most significant disturbing contribution to the system dynamics. Despite the numerous obstacles, the advantages associated with the selection of lower altitudes are relevant, especially with regards to Earth observation \citep{Crisp2020,Virgili-Llop2014} and communication applications \citep{Gavish1998}. The possibility of improving performance for comparable payload specifications or, alternatively, to reduce missions costs for a given performance profile has recently sparked the interest of the scientific community, as evidenced by the increasing number of studies investigating operations in VLEO \citep{CanasMunoz2020,Virgili-Llop2016,Fujita2009,Romano2020}. 

 However, as the harshness of the environment increases, the requirements imposed on the attitude determination and control system (ADCS) become more stringent. Momentum-based devices, such as reaction wheels (RWs) and control momentum gyroscopes (CMGs), are efficient means to compensate for cyclical disturbances, perform smooth re-orientation manoeuvring, and counteract the secular external perturbations affecting the satellite's dynamics and kinematics. However, in an environment where aerodynamic disturbance dominates, momentum wheels may be suceptible to faster saturation with an obvious impact on the possibility to sustain operations. 
If on one hand VLEOs represent a challenge for mission design, on the other hand they offer the unique opportunity to investigate the application of novel control schemes that take advantage of the thermospheric environment rather than trying to fight against it. As the aerodynamic torques are the prevalent source of external disturbance, investigating whether their magnitude and direction could be modified in order to provide a low cost means for momentum unloading appears reasonable. 
The enhanced aerodynamic torques experienced by satellites at these altitudes can also be employed to implement hybrid  active attitude control strategies. As many spacecraft are endowed with appendages extending from the bus, flat surfaces like solar panels can be used as aerodynamic actuators to support internal momentum devices while performing pointing control tasks. 
The combined employment of aerodynamic and conventional actuators may prove to be especially useful when actuator failure occurs \citep{Horri2012}, when size and volume restrictions are demanding, when at very low altitudes (<250 km) or during periods of intense solar activity the experienced aerodynamic torques may be too large for conventional actuators to handle the control task alone, or more generally when the requirement imposed on agility and accuracy are less onerous.

 The greatest challenge, when implementing active aerodynamic attitude control, is addressing the uncertainties that characterise the problem formulation. 
The majority of studies propose to achieve aerodynamic attitude control by selecting the angles of rotation of allocated aerodynamic control actuators \citep{Burns1992, Wang2014,Mostaza-Prieto2017,Gargasz2007,Llop2014,MP2017,CanasMunoz2020,Pande1979,Auret2011}. A simple and quick way of estimating the control torque is obtained by assuming small attitudes with regards to the incoming flow and limited ranges of deflection of the panels so that approximate linear laws can be used 
\citep{Burns1992,Gargasz2007,Wang2014,CanasMunoz2020,Auret2011,Mostaza-Prieto2017}. Alternative approaches see the implementation of an optimal control problem to find the time-varying position of the panels to minimise a desired cost function \citep{MP2017}, or the use of truncated Fourier series \citep{Johnson1971} to obtain an estimation of the aerodynamic torques. In order to privilege the demonstration of the control concept, some studies assume in first approximation time-invariant aerodynamic coefficients and neglect the small contribution due to the aerodynamic lift to the computation of the aerodynamic torques \citep{Burns1992, Wang2014, Pande1979}. For the computation of the aerodynamic coefficients, most investigations generally assume diffuse re-emission with complete \citep{Burns1992} or incomplete \citep{Mostaza-Prieto2017,MP2017,CanasMunoz2020} accommodation of the particles to the surface, or include some specular re-emission component by applying partial accommodation theory \citep{Gargasz2007,Auret2011}. Atmospheric density is modelled by means of exponential \citep{Gargasz2007,Auret2011}, sinusoid \citep{Burns1992}, and atmospheric models \citep{Mostaza-Prieto2017,CanasMunoz2020,MP2017}. 
Studies employing atmospheric models generally benefit from a more accurate description of other environmental parameters, such as the expected temperature at the selected altitude, atmospheric composition and molecular speed ratio, which are otherwise commonly neglected \citep{Gargasz2007,Auret2011,Johnson1971}.
Inclusion of thermospheric winds in the determination of the direction of impingement of the particles is usually ignored, with some exceptions \citep{CanasMunoz2020}. The applicability of aerodynamic control techniques have been evaluated against some of the uncertainties mentioned above in only a limited number of studies \cite{Pande1979,Mostaza-Prieto2017,MP2017}. 

This paper has the purpose of: 1) providing a comprehensive background on the challenges affecting aerodynamic torque modelling in VLEO;
 2) proposing an on-line algorithm for the determination of the control panels configuration that provides the closest match to a required input control signal; 3) testing the algorithm sensitivity to uncertainties and identifying critical conditions for operations for the case of some combined aerodynamic and RWs pointing manoeuvres. 
According to this, the paper is organised in the following way: \cref{aerodynamic_control_authority} discusses the sources of variability of the aerodynamic control authority in VLEO;
\cref{satellite_design} provides a brief description of the principal features of the satellite geometry assumed; \cref{satellite_dynamics_kinematics} defines the mathematical model for the satellite dynamics keeping into account the time-varying effects of the rotation of the appendages; \cref{AFAAC} is devoted to the design of the algorithm for selecting the angles of deflection of the aerodynamic actuators providing the desired control torque; validation under a limited set of assumptions and considering as many sources of uncertainty as possible, is addressed and discussed in \cref{Validation of the PCA algorithm and discussion}. Finally,  in \cref{MCanalysis}, the robustness of the algorithm is demonstrated by means of a Monte Carlo analysis  and potential limits of the controller are discussed to identify an optimal range of operations.

\section{Aerodynamic control authority in VLEO}\label{aerodynamic_control_authority}
Accurate prediction of the control authority achievable in VLEO is hindered by the complex mechanisms involved in the generation of the aerodynamic torques.
Because of this, works discussing means of exploitation of aerodynamic torques in VLEO generally rely on an apparatus of assumptions that are introduced with the scope of reducing the number of degrees of freedom involved. 
The parameters required to estimate the aerodynamic torques  ($\boldsymbol{T_{a}}$) can be explicitly or implicitly inferred from their well-known mathematical formulation:
\begin{equation}\label{aero_torque}
\boldsymbol{ T_{a}} = \boldsymbol r_{\boldsymbol{PO}} \times m\boldsymbol a_{\boldsymbol A} = \frac{1}{2}\rho \varv_{\mathrm{rel}}^2 S_{\mathrm{ref}} \ell_{\mathrm{ref}} \boldsymbol C_{\boldsymbol M}
\end{equation}
\noindent where $\boldsymbol r_{\boldsymbol {PO}}$ indicates the vector defining the distance between the CoP
and the CoM, $m$ is the satellite mass, $\boldsymbol a_{\boldsymbol A}$ is the aerodynamic acceleration vector, $\rho$ is the atmospheric density, $\varv_{\mathrm{rel}}$ is the magnitude of the satellite velocity with regards to incoming flow, $S_{\mathrm{ref}}$ and $\ell_{\mathrm{ref}}$ are the reference surface and length used to perform the computation and $\boldsymbol C_{\boldsymbol M} = \{C_{\phi}, C_{\varphi}, C_{\psi}\}$ is the vector of the three aerodynamic momentum coefficients in roll ($\phi$), pitch ($\varphi$) and yaw ($\psi$). As clearly stated by Eq. \ref{aero_torque}, alterations in the induced aerodynamic torques are expected with varying orbital/environmental conditions ($\frac{1}{2}\rho \varv_{\mathrm{rel}}^2$) and platform characteristics ($S_{\mathrm{ref}}\ell_{\mathrm{ref}} \boldsymbol C_{\boldsymbol M}$). 

\subsection{The role of the orbital and environmental conditions}\label{orbit_var}
\begin{figure}\begin{centering}
\includegraphics[width=1\textwidth]{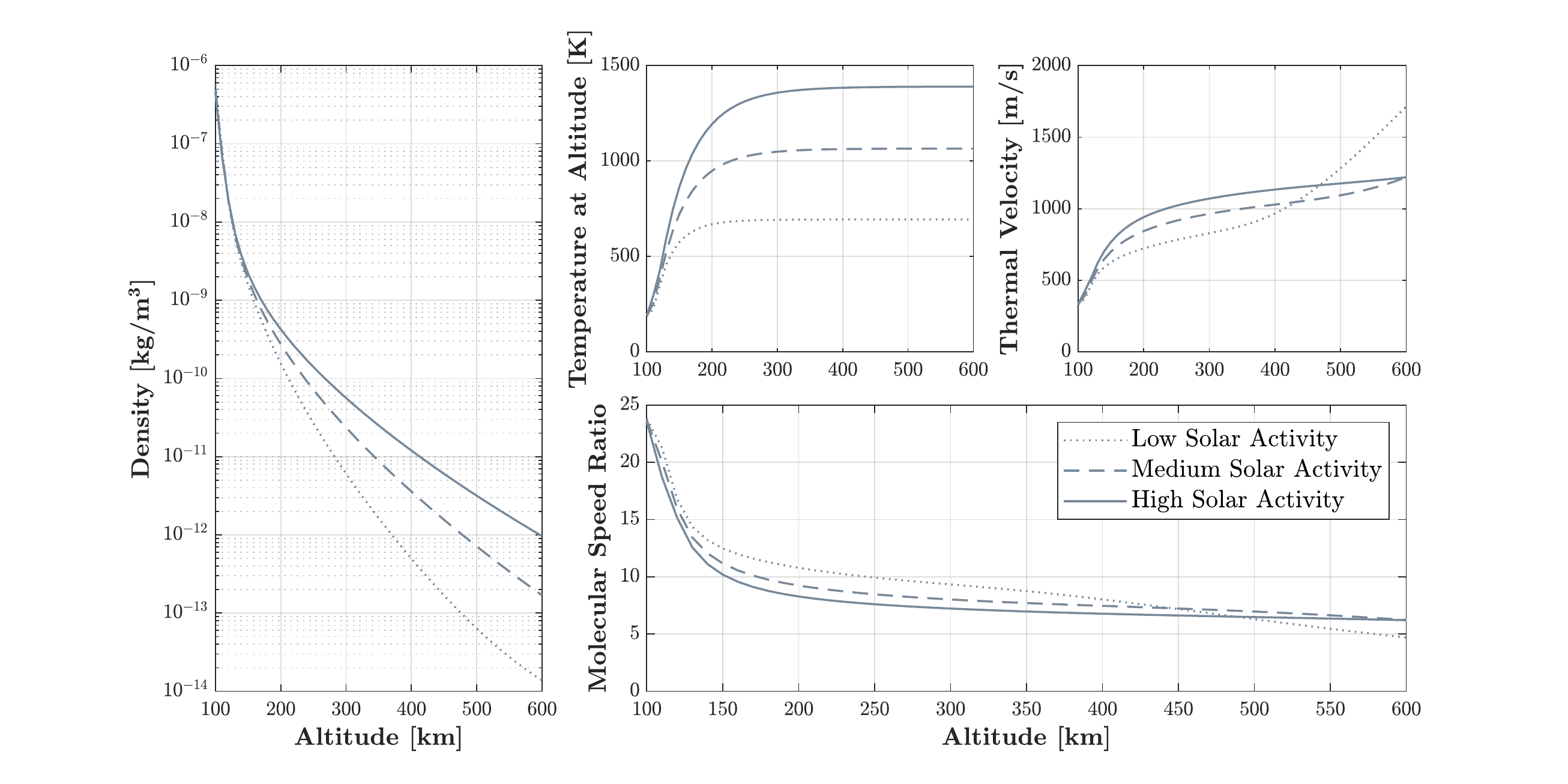}
\caption{Variation of environmental parameters with altitude and solar activity.}
\label{with_altitude}
\end{centering}
\end{figure}
The thermospheric environment mainly exerts its impact through the altitude-dependent variations of $\rho$ and $\varv_{rel}$, the components of the dynamic pressure:
\begin{equation}\label{dynamic_pressure}
q_{d} =  \frac{1}{2}\rho \varv_{rel}^2
\end{equation}
\noindent Less significant variations are observed in the molecular speed ratio ($s$), that defines the relative order of magnitude of the gas bulk velocity ($\varv_{b}$) and the most probable thermal speed of the particles ($\varv_{t}$):
\begin{equation} \label{molecular speed ratio}
s=\frac{\varv_{b}}{\varv_{t}}=\frac{\varv_{b}}{\sqrt{\frac{2RT_{alt}}{m_{m}}}}
\end{equation}
\noindent where $R = 8314 \text{\: J  kmol\textsuperscript{-1} K\textsuperscript{-1}}$ is the universal constant of gas, $m_{m}$ is the molecular mass of the atmospheric constituents and $T_{alt}$ is the temperature at the altitude considered in K. 
The thermospheric flow has a drifting Maxwellian velocity distribution characterised by $\varv_{t}\ll\varv_{b}$.
The effect of $\varv_{t}$ is thus generally negligible and can be ignored, unless aerodynamic coefficients are computed for grazing angles of incidence.
Although atmospheric density is mainly characterised by an exponential decay profile with altitude (Fig. \ref{with_altitude}, left), spatial and temporal fluctuations are observed in concomitance with alterations in the amount of thermal energy deposited in the lower thermosphere. The latter is subject to change with a number of physical processes, the most relevant being  the 11-year cyclical variation of solar activity, the day-to-night, the seasonal-latitudinal, and the annual and semi-annual variations \citep{Emmert2015}. 
\begin{figure}\begin{centering}
\includegraphics[width=1\textwidth]{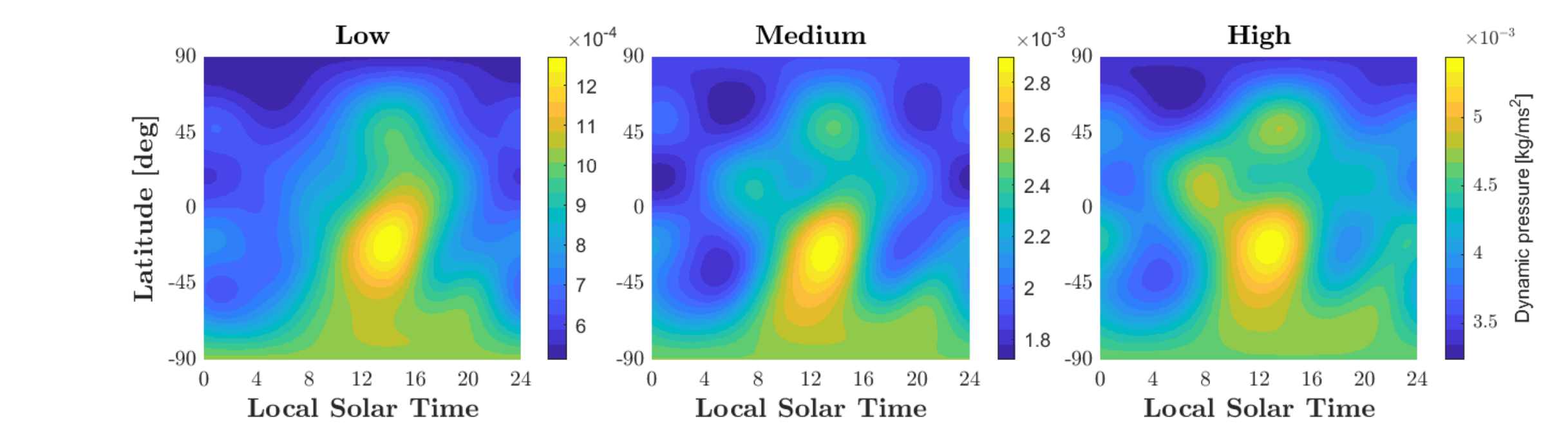}
\caption{Dynamic pressure variations with latitude, local solar time, and solar activity at 250 km.}
\label{dynamic_pressure}
\end{centering}
\end{figure}
 In Fig.\ref{with_altitude} the trends of the environmental parameters with altitude and solar activity for a reference longitude (${{\phi_{l}}= 0}\degree$) during low (1\textsuperscript{st} Jan 2020), medium (1\textsuperscript{st} Jan 2023), and high solar activity (1\textsuperscript{st} Jan 2026) is shown. The NRLMSISE-00 model \citep{Picone2002} and the solar and magnetic indices defined by ISO 1422:2013 \citep{ISO2013} are used.
 As a consequence of the thermospheric expansion caused by the increased energy absorption, the mean molecular weight of the atmospheric constituents at a given altitude increases during periods of high solar activity and an overall increase in neutral density is observed (Fig. \ref{with_altitude}, left). 
The effect of the particle thermal velocity at higher altitudes become more relevant as the solar activity attenuates (Fig. \ref{with_altitude}, top right), so that variations of the molecular speed ratio are more noticeable for quiet atmospheric conditions at altitudes > 400 km (Fig. \ref{with_altitude}, bottom). Altitudes < 150 km see abrupt variations of the environmental parameters with height and consequently bigger uncertainties in the estimation of the aerodynamic torques. The aerodynamic drag induced in these orbits is however so high that operations cannot be sustained for a meaningful duration without compensation.


Modifications in temperature, gradients of pressure, kinematic viscosity and molecules-to-atoms ratio further affects the dynamics of the upper atmosphere through the generation of atmospheric winds. 
As discussed in more detail in \cref{Relative velocity estimation}, the satellite inertial velocity at a certain altitude ($\boldsymbol{\varv_{I}}$)  and relative velocity with regards to the incoming flow ($\boldsymbol{\varv_{rel}}$) do not generally coincide, as the velocity relative to the flow also takes into account the contributions due to atmospheric corotation ($\boldsymbol{\varv_{rot}}$) and thermospheric winds ($\boldsymbol{\varv_w}$). 
While satellite inertial velocity and atmospheric co-rotation can be derived with substantial accuracy, the same cannot be said about the thermospheric winds component. 
As a consequence, a precise knowledge of the satellite attitude with regards to the incoming flow is not currently achievable. However, since the order of magnitude of the satellite inertial velocity is predominant, for the sole purpose of defining the aerodynamic control authority expected in VLEO, $\varv_{I} \simeq \varv_{rel}$ can be reasonably assumed. 
 This approximation was taken into account in Fig. \ref{dynamic_pressure}, which shows the local variations of the dynamic pressure predicted by the NRLMSISE-00 atmospheric model \citep{Picone2002} at 250 km for a range of latitudes, local solar times and solar conditions. 
Because of the Earth's rotation, the atmosphere warming is not homogeneous and a bulge, corresponding to a maximum in $\rho$, is clearly visible in all the three maps. Whilst the core appears to be centered at local solar time $\simeq$14$-$15, its peripheral extension on both latitude and local solar time and its magnitude vary significantly according to solar cycle progression and the Sun's declination.
\subsection{The role of the variables related to the platform design}\label{engineering variables}
Some control over the generated aerodynamic torques can be achieved through a proper selection of the satellite geometry and the materials employed for the external surfaces.
In VLEO, despite the higher number density of constituents, 
the non-dimensional Knudsen number ($K_{n}$) is large enough so that $K_{n} \rightarrow \infty$ can be reasonably assumed. In these conditions, the flow is characterised by a high degree of rarefaction and it is generally referred to as a \textit{free molecular flow} (FMF). The FMF regime
 is characterised by the predominance of surface-particles collisions over inter-particles collisions. As a consequence, the leading mechanism involved in the generation of the aerodynamic torques in VLEO is identifiable with the thermal energy and momentum exchange between the incident atmospheric particles and the external surfaces. 
A number of gas-surface interaction (GSI) models have been developed with the scope of providing a mathematical formulation for the aerodynamic coefficients and the exchange with the surfaces \citep{Sentman1961,Moe2005,Schamberg1959,Cook1966,Schaaf1958,Maxwell1890,Cercignani1971}. 
Reviews of some well-known GSI models in relation to the orbital aerodynamics problem are available \citep{MostazaPrieto2014b, Livadiotti2020}. 
However, at present a simple GSI model capable of capturing different interaction scenario with accuracy is not available. 
Due to the high degree of contamination by atomic oxygen (AO) adsorption \citep{Banks2004}, common materials used on spacecraft typically show diffuse re-emission patterns corresponding to high energy and momentum exchange. 
However, gas-beam experimental results show that quasi-specular lobular re-emission patterns, characterised by reduced accommodation to the surface, are achievable when smooth and clean surfaces are employed \citep{Goodman1971}. Few studies showing interest towards a novel generation of aerodynamic materials resistant to AO adsorption are already present in literature \citep{Murray2017,CrispNicholasH2020}. If their scattering characteristic can be demonstrated to be consistent over time, there might be an attractive possibility to employ higher-performing aerodynamic materials for on-orbit applications.
\begin{figure}
\begin{centering}
\def\svgwidth{160mm}
\includegraphics[width=1\textwidth]{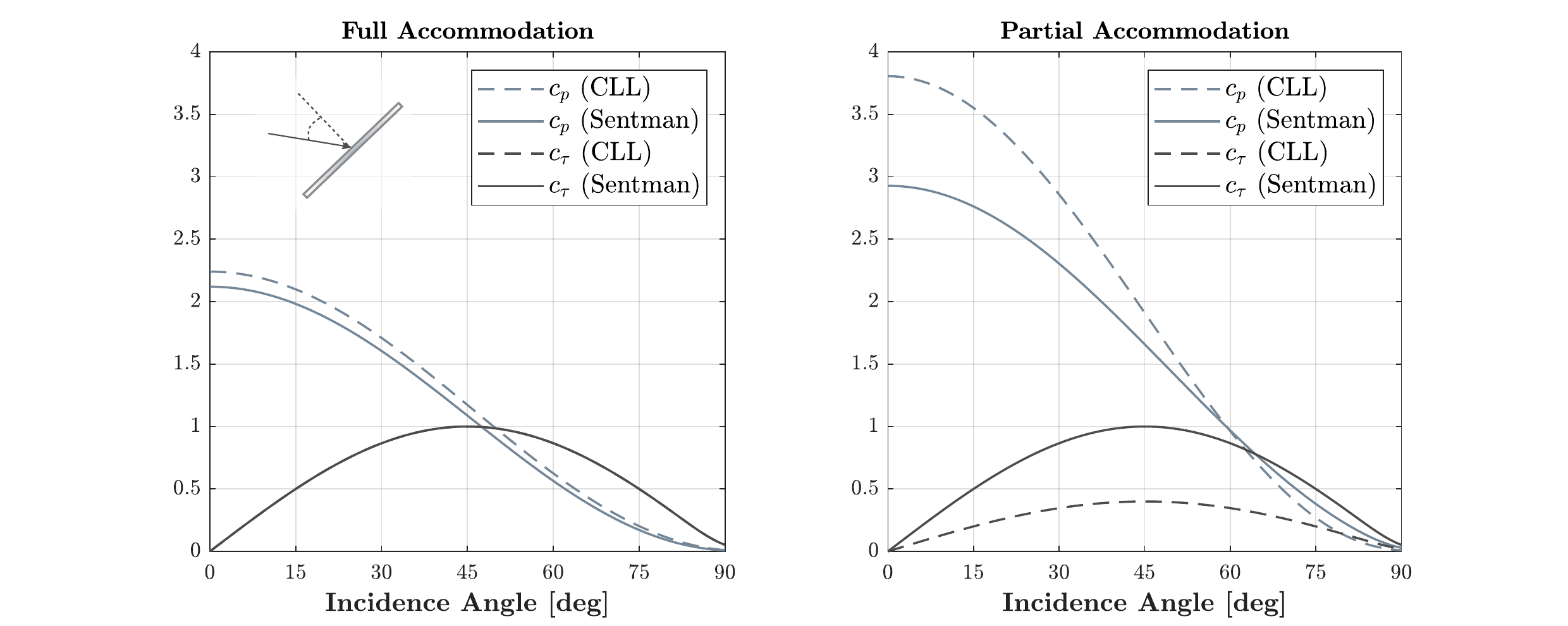}
\caption{Variation of $ {c_{p}}$ and ${c_{\boldsymbol\tau}}$ on a flat plate with the re-emission characteristic (GSI model), the particles degree of accommodation, and the incident angle. }
\label{models_accommodation}
\par\end{centering}
\end{figure}
Fig. \ref{models_accommodation} compares the aerodynamic performance of a flat plate for varying re-emission profiles, incidence angles and incident particle accommodation.
To perform this analysis, Sentman's model \citep{Sentman1961} was employed to model diffuse re-emission, while the Schaaf and Chambre analytical equations modified by Walker et al. \citep{Walker2014} according to the Cercignani-Lampis-Lord (CLL) model assumptions were used to simulate the performance of quasi-specular materials. Uncertainties in the estimation of the aerodynamic coefficients is given by the normal pressure ($c_{p}$) and the shear stress ($c_{\tau}$) coefficient components. Substantial qualitative and quantitative agreement between the models is observed when the incident particles attain complete accommodation with the surfaces (Fig. \ref{models_accommodation}, left), as deviations mainly lie in the normal pressure component. 
Conversely, the predicted values of $c_{p}$ and $c_{\tau}$ change significantly with the model and the re-emission mechanism assumed in the presence of partial accommodation (Fig. \ref{models_accommodation}, right).
Higher aerodynamic coefficients, and thus potentially higher aerodynamic control authority, is achievable for quasi-specular re-emissions and reduced particle accommodation.
For intermediate scenarios, an imprecise knowledge of the particles accommodation to the surface translates to a source of uncertainty in aerodynamic modelling.

\begin{figure}
\begin{centering}
\def\svgwidth{160mm}
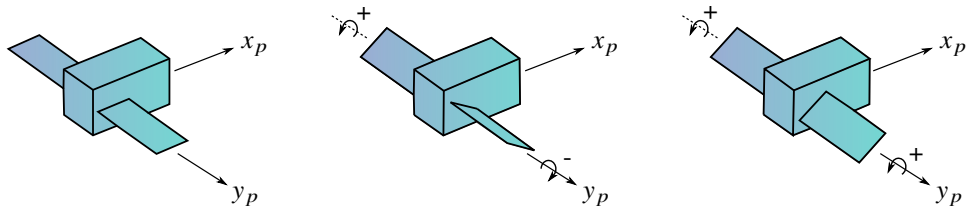
\caption{Generic 2U CubeSat geometry with minimum drag (left), symmetrical counter-rotated (middle), and symmetrical co-rotated (right) configurations.}
\label{CubeSat}
\par\end{centering}
\end{figure}

\begin{figure}\begin{centering}
\includegraphics[width=1\textwidth]{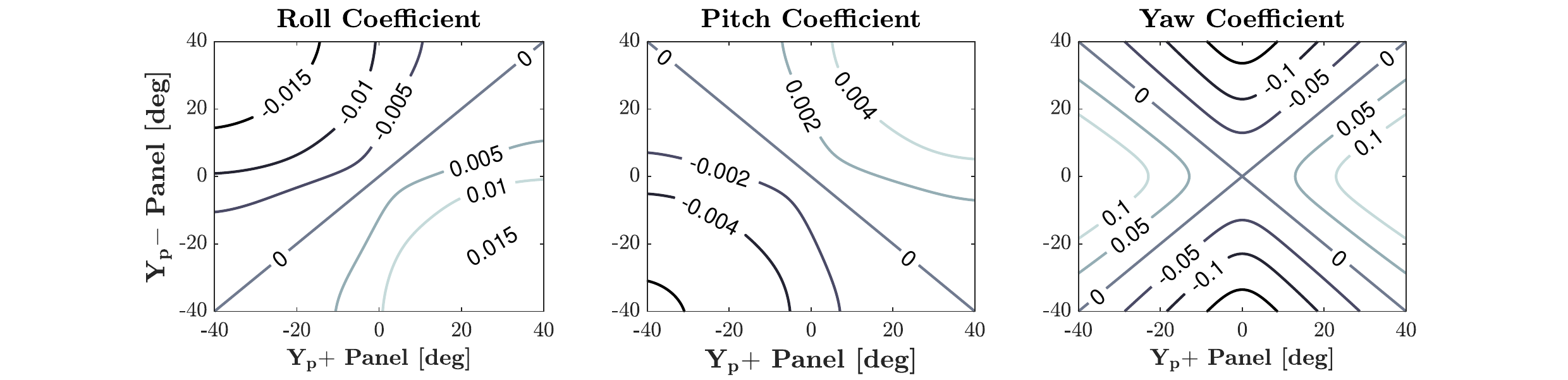}
\caption{Variation of the aerodynamic coefficients in roll, pitch and yaw with the relative angles of deflection of two aerodynamic panels extending along the side of a cuboid satellite.}
\label{geometry_impact}
\end{centering}
\end{figure}

\begin{figure}\begin{centering}
\includegraphics[width=0.8\textwidth]{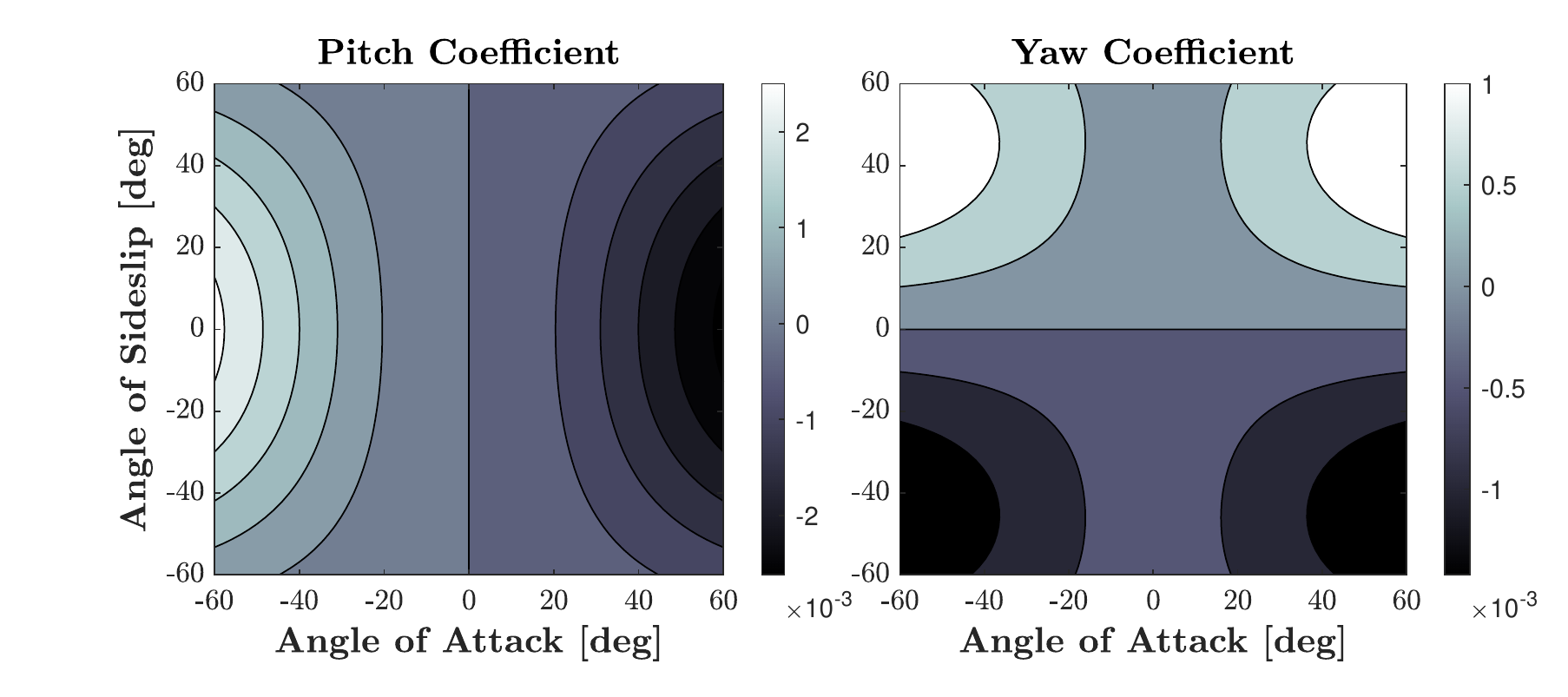}
\caption{Variation of the aerodynamic coefficients in pitch (left) and yaw (right) with attitude.}
\label{attitude_impact}
\end{centering}
\end{figure}

As revealed by Eq. \ref{aero_torque} and by Fig. \ref{models_accommodation} , the aerodynamic control capability is also susceptible to the geometric characteristics of the platform.
For a given attitude with regards to the flow, the induced aerodynamic coefficients vary with the area exposed to the flow and with the angle at which the atmospheric particles impact the surfaces. Alterations in both characteristics are obtainable by moving or rotating some designated control actuators. 
The effect of external geometry alteration on the induced non dimensional aerodynamic coefficients is shown in Fig. \ref{geometry_impact}, for a satellite with a cuboid bus and two aerodynamic panels located at the rear of the CoM (Fig. \ref{CubeSat}). 
Variation in the induced pitch and yaw torques can be easily achieved by introducing some asymmetries in the satellite external configuration
so that corotation of the panels can be commanded for this purpose (Fig. \ref{geometry_impact}, middle and right). Conversely, variations in the roll coefficient require counter-rotation of the panels, with better results achievable when symmetrical configurations are selected (Fig. \ref{geometry_impact}, left).
For a given configuration of the actuators, variations in the induced aerodynamic coefficients are also expected to occur with attitude, as evidenced by the maps for the pitch and yaw coefficients of Fig. \ref{attitude_impact} that refer to the 2U satellite in its minimum drag configuration (Fig. \ref{CubeSat}, left), the total area of the panels, and the relative distance between the CoP and the CoM.

Due to the flow conditions, for a given surface exposed to the flow, discrepancies between the expected and the effectively induced aerodynamic torques may arise from panel shadowing: this phenomenon occurs when the satellite attitude with regards to the flow is such that ram surfaces shield some portions of the downstream surfaces, preventing their interaction with the atmospheric particles. If shadowing affects control panels, a reduced flow interaction translates to an inferior achievable control authority. This effect is especially important when concave geometries in non-equatorial orbits are considered and the magnitude of thermospheric winds is relevant.
 In general, a convex design with long panels largely
prevents surface shielding and multiple particles interactions from occurring. 
Under these circumstances,
this phenomenon eventually involves just a very limited portion of the exposed surfaces. 

The introduction of geometric considerations in VLEO is fundamental not only to increase the aerodynamic control authority, but also to minimise the induced aerodynamic drag. Because of its dissipative nature, it represents the primary drawback affecting applications in this altitude range. The use of movable control surfaces to perform attitude control has an impact on the induced aerodynamic drag as a consequence of the generally increased projected area exposed, during manoeuvring, to the flow. Some shaping criteria, however, can be employed especially with regards to the satellite main body \citep{Walsh2017,Park2014}: while the panel configuration can be modified so that aerodynamic forces and torques can be usefully employed, the aerodynamic contribution due to the satellite main body represents a source of attitude and orbital disturbance for which minimisation is desirable.
\section{Satellite geometry and aerodynamic design}\label{satellite_design}
\begin{figure}
\begin{centering}
\def\svgwidth{140mm}
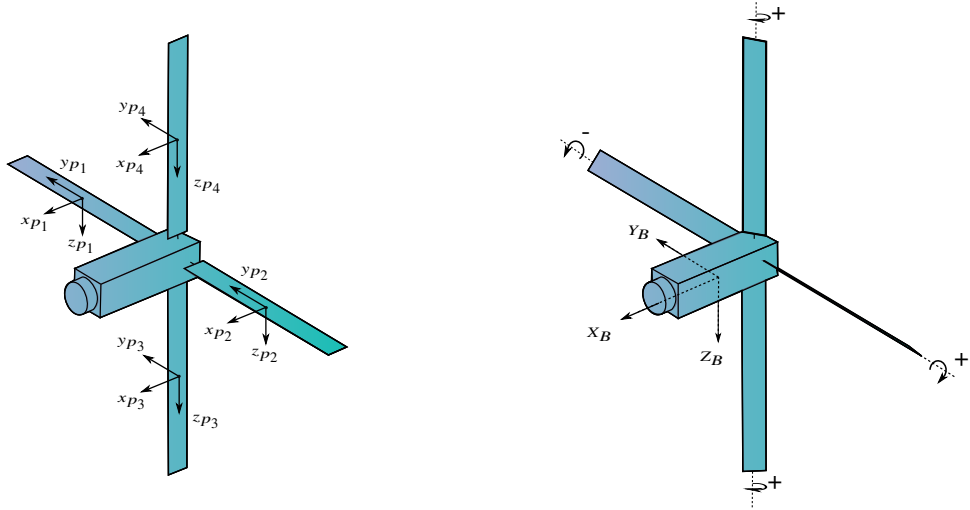
\caption{SOAR's geometry for two representative configurations. Left: nominal low-drag configuration. Right: arbitrary panels configuration. }
\label{SOAR_CubeSat}
\par\end{centering}
\end{figure}

The feathered geometry of SOAR \citep{Crisp2021} is selected as reference for the problem formulation and the validation of the results. The essential geometric characteristics of this satellite are shown in Fig. \ref{SOAR_CubeSat}: SOAR is a 3U CubeSat characterised by four rotating panels mounted at rear of the satellite main body. 
 Enhanced control authority in three-axes is achieved by setting the incidence angle of each panel with regards to the incoming flow independently: co-rotated and counter-rotated configurations of the vertical and horizontal panels can thus be achieved by rotating the selected appendages about their longitudinal axes (Fig. \ref{SOAR_CubeSat}, right). Four reference frames are accordingly defined (x\textsubscript{p\textsubscript{i}}y\textsubscript{p\textsubscript{i}}z\textsubscript{p\textsubscript{i}} for i = 1,...,4) to describe the relative motion of each panel with regards to the moving satellite body reference frame X\textsubscript{B}Y\textsubscript{B}Z\textsubscript{B}, which is assumed to be centered at the satellite composite CoM. Their orientation in the satellite nominal low-drag configuration is shown in Fig. \ref{SOAR_CubeSat}. The origin of each panel reference frame is located  at the centre of mass of the corresponding panel, which is assumed perfectly symmetric about its rotational axis and uniform in its mass distribution. The X\textsubscript{B}Y\textsubscript{B}Z\textsubscript{B} reference frame and the panels reference frames are aligned when the rotation angles of the four panels is set to zero, i.e. when the panels are in the nominal low drag configuration. For any other configuration, the orientation of each panel reference frame with regards to X\textsubscript{B}Y\textsubscript{B}Z\textsubscript{B} can be described by the associated direction cosine matrix. Positive and negative rotations of the aerodynamic surfaces are defined according the X\textsubscript{B}Y\textsubscript{B}Z\textsubscript{B} reference frame description following the right hand rule and they are consistent among the aerodynamic surfaces employed (Fig. \ref{SOAR_CubeSat}, right).
\section{Satellite dynamics with rotating appendages}\label{satellite_dynamics_kinematics}
When rotating aerodynamic actuators are employed for control purposes, the satellite dynamics about the center of mass is not only affected by the presence of the internal momentum devices (if present) but also by the rotational dynamics of these appendages. According to the principle of conservation of the angular momentum, an equivalence can be established between the inertially referenced rate of change of the angular momentum of a system about its centre of mass ($\boldsymbol{H^{I}_{tot}}$) and the external torques acting on it ($\boldsymbol{T_{e}}$).
Applying Euler's momentum formulation:
\begin{equation}\label{dot_H_tot_I}
\boldsymbol{\dot H^{I}_{tot}}=\boldsymbol{\dot H^{B}_{tot}} + \boldsymbol{\omega^{I}_{B}}\times \boldsymbol{H^{B}_{tot}}
\end{equation}
so that:
\begin{equation}\label{dot_H_tot_B}
\boldsymbol{\dot H^{B}_{tot}} = \boldsymbol{T_{e}} - \boldsymbol{\omega^{I}_{B}}\times \boldsymbol{H^{B}_{tot}}
\end{equation}
In Eqs. \ref{dot_H_tot_I} and  \ref{dot_H_tot_B}, $\boldsymbol{T_{e}}$ generally includes the contribution due to aerodynamic torques ($\boldsymbol{T_{a}}$), disturbance torques ($\boldsymbol{T_{d}}$) and the control torques provided by the actuators, $\boldsymbol{H^{B}_{tot}}$  is the total angular momentum in body axes, and $\boldsymbol{\omega^{I}_{B}}$ are the inertially referenced body angular rates. For the system considered in this paper, $\boldsymbol{H^{B}_{tot}}$ is given by the sum of the angular momentum of the total satellite system ($\boldsymbol{H^{B}_{s}}$), the angular momentum of each rotating surface about its center of mass ($\boldsymbol{H^{B}_{p}}$) and the angular momentum of the RWs about their axis of rotation ($\boldsymbol{H^{B}_{w}}$), all represented in the satellite body frame:
 \begin{equation}\label{H_B_tot}
 \begin{split}
\boldsymbol{H^{B}_{tot}} =  \boldsymbol{H^{B}_{s}} + \boldsymbol{H^{B}_{p}} + \boldsymbol{H^{B}_{w}} =&\:  \boldsymbol{J_{s}}\boldsymbol{\omega^{I}_{B}} +
\sum_{i = 1}^{n_p}A^{p,i}_{B} \boldsymbol{J_{p,i}}\boldsymbol{\omega_{p,i}} +\sum_{q = 1}^{n_w}\boldsymbol{J_{w,q}}\left(\boldsymbol{a^{w,q}_{B}}\cdot\boldsymbol{\omega^{I}_{B}} + \boldsymbol{\omega_{w,q}}\right)\boldsymbol{a^{w,q}_{B}}
 \end{split}
\end{equation}
where $\boldsymbol{J_{s}}$, $\boldsymbol{J_{p}}$ and $\boldsymbol{J_{w}}$  are respectively the inertia tensors of the total satellite, the panels, and the RWs, $\boldsymbol{\omega_{p}}$ and $\boldsymbol{\omega_{w}}$ indicate the angular rate of the panels and of the wheels about their spin axis, $A^{p,i}_{B}$ is the rotation matrix from the i-th appendage to the body reference frame, and $\boldsymbol{a^{w,q}_{B}}$ is the vector that defines the orientation of the q-th wheel in body axes. Computing the first time derivative of Eq. \ref{H_B_tot} relatively to the body reference frame, it is possible to get:
\begin{equation}\label{dot_H_tot_B_I}
\begin{split}
\boldsymbol{\dot H^{B}_{tot}} =&\: \boldsymbol{J_{s}}\boldsymbol{\dot\omega^{I}_{B}} + \boldsymbol{\dot J_{s}}\boldsymbol{\omega^{I}_{B}} +
\sum_{i = 1}^{n_p}A^{p,i}_{B} \boldsymbol{J_{p,i}}\left(\boldsymbol{\dot\omega_{p,i}}\right)_{B} + \sum_{i = 1}^{n_p}A^{p,i}_{B} \left(\boldsymbol{\dot J_{p,i}}\right)_{B}\boldsymbol{\omega_{p,i}} +
\sum_{q = 1}^{n_w}\boldsymbol{J_{w,q}}\left(\boldsymbol{a^{w,q}_{B}}\cdot\boldsymbol{\dot\omega^{I}_{B}} + \boldsymbol{\dot\omega_{w,q}}\right)\boldsymbol{a^{w,q}_{B}}
\end{split}
\end{equation}
Assuming that the time variation of the total satellite inertia matrix is only due to the rotation of the aerodynamic surfaces with regards to the satellite main body \citep{Rheinfurth1985}:
\begin{equation}
\boldsymbol{\dot J_{s}} = \sum_{i = 1}^{n_p}A^{p,i}_{B}\left(\boldsymbol{\dot J_{p,i}}\right)_{B}{A^{B}_{p,i}}
\end{equation}
where $A^{B}_{p,i} = {A^{p,i}_{B}}^{T}$. According to this, Eq. \ref{dot_H_tot_B_I} can be re-arranged in the following way:
\begin{equation}\label{dot_H_tot_B_II}
\begin{split}
\boldsymbol{\dot H^{B}_{tot}} =&\: \boldsymbol{J_{s}}\boldsymbol{\dot\omega^{I}_{B}} +
\boldsymbol{\dot J_{s}}\left(\boldsymbol{\omega^{I}_{B}}  + \sum_{i = 1}^{n_p}A^{p,i}_{B} \boldsymbol{\omega_{p,i}}\right) +
\sum_{i = 1}^{n_p}A^{p,i}_{B} \boldsymbol{J_{p,i}}\left(\boldsymbol{\dot\omega_{p,i}}\right)_{B}+ \sum_{q = 1}^{n_w}\boldsymbol{J_{w,q}}\left(\boldsymbol{a^{w,q}_{B}}\cdot\boldsymbol{\dot\omega^{I}_{B}} + \boldsymbol{\dot\omega_{w,q}}\right)\boldsymbol{a^{w,q}_{B}}
\end{split}
\end{equation}
Recognising that:
\begin{equation}\label{RW_torque}
\boldsymbol{\dot H^{B}_{w}} = \sum_{q = 1}^{n_w}\boldsymbol{J_{w,q}}\left(\boldsymbol{a^{w,q}_{B}}\cdot\boldsymbol{\dot\omega^{I}_{B}} + \boldsymbol{\dot\omega_{w,q}}\right)\boldsymbol{a^{w,q}_{B}}
\end{equation}
and substituting Eq. \ref{H_B_tot}, \ref{dot_H_tot_B_II} and \ref{RW_torque} in Eq. \ref{dot_H_tot_B}, it is possible to get:
\begin{equation}\label{omega_B_I}
\begin{split}
\boldsymbol{\dot\omega^{I}_{B}} = &\: \boldsymbol{J_{s}}^{-1}\left[\boldsymbol{T_{d}} + \boldsymbol{T_{a}} -
\boldsymbol{\dot J_{s}}\left(\boldsymbol{\omega^{I}_{B}}  +
\sum_{i = 1}^{n_p}A^{p,i}_{B} \boldsymbol{\omega_{p,i}}\right) -
\sum_{i = 1}^{n_p}A^{p,i}_{B} \boldsymbol{J_{p,i}}\boldsymbol{\dot\omega_{p,i}} -
\boldsymbol{\omega^{I}_{B}}\times\left(\boldsymbol{J_{s}}\boldsymbol{\omega^{I}_{B}} + 
\sum_{i = 1}^{n_p}A^{p,i}_{B} \boldsymbol{J_{p,i}}\boldsymbol{\omega_{p,i}} +
 \boldsymbol{H^{B}_{w}}\right) -
\boldsymbol{\dot H^{B}_{w}} \right]
\end{split}
\end{equation}
Removing the contributions due to the appendages rotations in Eq. \ref{omega_B_I}, the well known formulation of the rigid-body rotational equation for a satellite equipped with reaction wheels is obtained. For the specific case of the geometry considered, 
the matrix $A^{B}_{p,i} $, which describes the transformation to be applied from the body axes to the selected panels reference frame, assumes the form $A^{B}_{p,1/2}$ for the horizontal panels and $A^{B}_{p,3/4}$ for the vertical ones:
\begin{equation}\label{dcm: panels rotation}
A^{B}_{p,1/2} =
\begin{bmatrix}
cos(\vartheta_{p,1/2})&0&-sin(\vartheta_{p,1/2})\\
0&1&0\\
sin(\vartheta_{p,1/2})&0&cos(\vartheta_{p,1/2})
\end{bmatrix}\quad\quad
A^{B}_{p,3/4} =
\begin{bmatrix}
cos(\vartheta_{p,3/4})&sin(\vartheta_{p,3/4})&0\\
-sin(\vartheta_{p,3/4})&cos(\vartheta_{p,3/4})&0\\
0&0&1
\end{bmatrix}
\end{equation}
where $\vartheta_{p,1/2}$ and $\vartheta_{p,3/4}$ indicate the angle of rotation of the horizontal and vertical panels, respectively. More significant perturbations are introduced in the system when both the vertical and horizontal panels are symmetrically corotated and high angular velocities $\boldsymbol{\omega_{p,i}}$ of the panels are used. Disturbances cancel for the selection of symmetrical counter-rotated configurations of pairs of opposing panels.

\section{On-line algorithm for active aerodynamic attitude control}\label{AFAAC}
\begin{figure}[hbt!]
\begin{centering}
\includegraphics[width=1\textwidth]{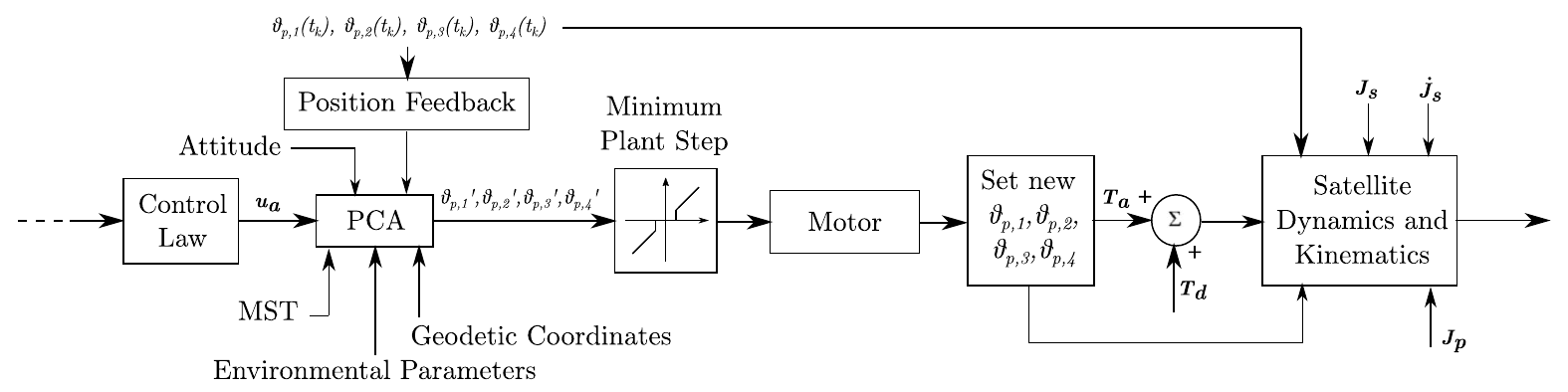}
\caption{General representation of the framework for the implementation of the PCA algorithm.}
\label{fig: PCA generic scheme}
\par\end{centering}
\end{figure}

In order to perform aerodynamic active attitude control, the commanded torque computed by a selected control law ($\boldsymbol{u_{a}}\approx \boldsymbol{T_{a}}$) at a given time step $t_{k}$ needs to be provided in input to an algorithm that accordingly selects the corresponding angles of deflection of the aerodynamic actuators. The decision process of the algorithm proposed in this study - here referred to with the acronym PCA (panel configuration algorithm) - is based on the on-line computation of the expected aerodynamic torques for the current satellite attitude. 
A qualitative and general representation of the flow of operations that, from the control signal produced by a generic control law, leads to the selection of the angles of deflection of the panels is provided in Fig. \ref{fig: PCA generic scheme}. A detailed description is given in the following.
\subsection{Atmospheric density modelling}
In order to take into account the discrepancies observed with altitude, solar activity, geographic location and diurnal variations, the approach used in this study relies on the linear interpolation of some reference values of $\rho$ contained in a lookup table. For given solar weather conditions, these values are computed by using the NRLMSISE-00 atmospheric model \citep{Picone2002} for a limited range of variation of VLEO altitudes, latitudes and local solar times. To limit the size of the lookup table within acceptable dimensions, the solar activity is fixed and selected according to the expected solar weather conditions at a reference epoch. 

Linear interpolation of the atmospheric density values in the lookup table requires the estimation of the geodetic latitude ($\phi_{gd}$), altitude ($h$) and local solar time at the time step $t_{k} = t_{0} + kt_{s}$, where $t_{0}$ is the initial epoch expressed in ephemeris seconds past J2000 and $t_{s}$ is the sampling time. The geodetic latitude and the altitude are computed from the ITRF93 referenced satellite position vector according to the algorithm proposed in \citep{AeroTool}, which was modified to use the Earth's equatorial radius ($r_{e} = 6378.2064$ km) and flattening coefficient ($f=1/294.9787$) of the Clark66 spheroid. Since the accuracy requirement imposed on the calculation of the local solar time is moderate, an approximate but quick implementation is preferred. If the small differences between UT1 and barycentrical dynamical time are neglected, the corresponding number of Julian centuries referred to a given $t_{k}$ are approximately equivalent ($T_{UT1}\simeq T_{TDB}$). The Sun's mean longitude ($\lambda_{M_\odot}$), mean anomaly ($M_{\odot}$) and ecliptic longitude ($\lambda_{E_{\odot}}$) can be accordingly computed as \cite[pp. 277--281]{Vallado2013}:
\begin{subequations}
\begin{gather}
\lambda_{M_{\odot}} = 280.460\degree + 36000.771\; T_{UT1}\\
M_{\odot} \simeq 357.5291092\degree +35999.05034\; T_{UT1}\\
\lambda_{E_{\odot}} = \lambda_{M_{\odot}} + 1.914666471\degree\sin(M_{\odot}) + 0.019994643\sin(2M_{\odot})
\end{gather}
\end{subequations}
Once the obliquity of the ecliptic ($\epsilon$) has been determined:
\begin{equation}
\epsilon \simeq 23.439291\degree-0.0130042\;T_{UT1}
\end{equation}
the Sun's position vector ($\boldsymbol{r_{\odot}}$) can be derived using \cite[pp. 277--281]{Vallado2013}:
\begin{subequations}
\begin{gather}\label{Sun_v}
r_{\odot} = 1.000140612 -0.016708617\cos(M_{\odot})-0.000139589\cos(2M_{\odot})\\
\label{Sun_vector}
\boldsymbol{r_{\odot}} =\left[r_{\odot}\cos(\lambda_{E_{\odot}}),r_{\odot}\cos(\epsilon)\sin(\lambda_{E_{\odot}}),r_{\odot}\sin(\epsilon)\sin(\lambda_{E_{\odot}})\right]^{T} \textrm{AU}.
\end{gather}
\end{subequations}
After converting \ref{Sun_vector} to the proper unit of measurement, the satellite inertial position vector ($\boldsymbol{r_{I}}$) and the Sun's position vector ($\boldsymbol{r_{\odot}}$) are used to compute the local hour angle \cite[p. 1002]{Vallado2013}:
\begin{equation}
LHA_{\odot} = \frac{180\degree}{\pi}\left[\frac{r_{\odot,x}r_{I,y}-r_{\odot,y}r_{I,x}}{\vert r_{\odot,x}r_{I,y}-r_{\odot,y}r_{I,x}\vert}\arccos\left(\frac{r_{\odot,x}r_{I,x}+r_{\odot,y}r_{I,y}}{\sqrt{r_{\odot,x}^{2} + r_{\odot,y}^{2} }\sqrt{r_{I,x}^{2} + r_{I,y}^{2} }}\right)\right]
\end{equation}
The mean local solar time (MST) is finally computed as the difference between the local apparent solar time ($LAST$) and the equation of time ($E_{Q}$), where \cite[p. 178]{Vallado2013}:
\begin{subequations}
\begin{gather}
LAST = LHA_{\odot} + 180\degree\\
E_{Q} = -1.914666471\degree\sin(M_{\odot}) -0.019994643\sin(2M_{\odot}) + 2.466\sin(2\lambda_{E_{\odot}}) -0.0053\sin(4\lambda_{E_{\odot}})
\end{gather}
\end{subequations}
\subsection{Relative velocity estimation}\label{Relative velocity estimation}
As discussed in  \cref{aerodynamic_control_authority}, the induced aerodynamic coefficients have a dependence on the satellite attitude with regards to the incoming flow. This can be deduced from the expression of $\boldsymbol{\varv_{rel}}$ in the J2000 reference frame:
\begin{equation}\label{v_rel}
\boldsymbol{\varv_{rel}} = \boldsymbol{\varv_{I}} + \boldsymbol{\varv_{rot}} + \boldsymbol{\varv_{w}} = \boldsymbol{\varv_{I}} + \left(\boldsymbol{\omega_{\oplus}}\,\times\boldsymbol{r_{I}}\right) + \boldsymbol{\varv_{w}}
\end{equation}
 In order to simulate an imperfect knowledge of the incoming flow direction (Section \ref{orbit_var}), the contribution due to $\boldsymbol{\varv_{w}}$ in Eq. \ref{v_rel} is not considered by the PCA. The vector $\boldsymbol{\varv_{rel}}$ derived accordingly is then used to determine, at each time step $t_{k}$, the rotation matrix from body-axes to wind-axes ($A^{B}_{F}$) and the corresponding angle of attack ($\alpha_{k}$) and sideslip ($\beta_{k}$):
\begin{equation}\label{eq:attack_sideslip}
\alpha_{k} = \arcsin{\left(-A^{B}_{F}\left(3,1\right)\right)} \quad\text{and}\quad 
\beta_{k} = \arcsin\left({A^{B}_{F}\left(1,2\right)}\right)
\end{equation}
\subsection{Selecting the configuration of the panels}\label{SCP}
Once $q_{d}$ and the satellite attitude with regards to the impinging flow ($A^{F}_{B} = {A^{B}_{F}}^{T}$) have been determined, the PCA selects the panel configuration providing the closest match between the induced aerodynamic torques $\boldsymbol{T_{a}}$ (Eq. \ref{aero_torque}) and the input control signal $\boldsymbol{u_{a}(t_{k})}$. In order to compute the dimensional aerodynamic coefficients $\boldsymbol{C_{M_{d}}}=S_{ref}\ell_{ref}\boldsymbol{C_{M}}$ in Eq. \ref{aero_torque}, a quick online computation technique is utilised. This algorithm is based on ADBSat \citep{Mostaza-Prieto2017,Sinpetru2020}, a panel method for aerodynamic performance computation developed at the University of Manchester. The original ADBSat software, envisaged as an off-line tool to determine the aerodynamic characteristics of a geometry immersed in FMF, was modified to provide online computation of the dimensional aerodynamic coefficients for multiple control panel permutations and varying re-emission characteristics under the computational speed requirement imposed by the stability of a discrete-time control loop with a sampling time $t_{s} = 1\text s$. 

The PCA firstly reads the cartesian coordinates of a simplified triangular mesh that describes the geometry of the satellite in its nominal low drag configuration $P_{N} = [0\degree,0\degree,0\degree,0\degree]$. 
The total number of elements in the mesh ($M_{tot}$) is limited not only to reduce the computational effort, but also to contain the size of the input data used by the algorithm. 
A geometric criteria, based on the characteristic dimensions of the satellite, is then used to identify the triangular surface elements in the mesh belonging to the $p_{1}$, $p_{2}$, $p_{3}$ and $p_{4}$ panels 
from those that belong to the satellite main body:
\begin{subequations}
\begin{gather}
M_{j}\in p_{1}\Leftrightarrow\forall\,\,Y_{B,m}\in M_{j},\, Y_{B,m}\ge \left(w_{mb}/2 - Y_{CoM} + l_{s}\right)\\
M_{j}\in p_{2}\Leftrightarrow\forall\,\,Y_{B,m}\in M_{j},\, Y_{B,m} \le -\left(w_{mb}/2+ Y_{CoM} + l_{s}\right)\\
M_{j}\in p_{3}\Leftrightarrow\forall\,\,Z_{B,m}\in M_{j}, \,Z_{B,m} \ge \left(w_{mb}/2- Z_{CoM} + l_{s}\right)\\
M_{j}\in p_{4}\Leftrightarrow\forall\,\,Z_{B,m}\in M_{j},\,Z_{B,m}\le -\left(w_{mb}/2+ Z_{CoM} + l_{s}\right)
\end{gather}
\end{subequations}
where $M_{j}$ is the j-th element of the triangular mesh inspected, $Y_{B,m}$ and $Z_{B,m}$ with $m =1,2,3$ are the body-referenced coordinates of the m-th vertex that constitutes the $M_{j}$ triangular mesh element considered, $Y_{CoM}$ and $Z_{CoM}$ are the body-referenced coordinates of the CoM measured from the geometric centre of the satellite, $w_{mb}$ is the width of the satellite main body and $l_{s}$ is the length of the support by which each panel is connected to the main body. To simulate the aerodynamic performance of different materials, a thermal accommodation coefficient ($\alpha_{T,j}$), describing the extent to which the particles achieve thermal accommodation with the surface, is associated to the elements identified in the mesh:
\begin{equation}\label{alpha_accom}
\alpha_{T,j} = \frac{T_{i}-T_{r,j}}{T_{i}-T_{w,j}}
\end{equation}
where $T_{i}$ and $T_{r,j}$ are respectively the kinetic temperatures of the incident and re-emitted particles and $T_{w,j} = T_{w} = 300 \text{ K} = \text{const}$ is the averages surface temperature. Variations of the surface temperature, albeit present, are not modelled, as the associated variations of the thermal accommodation coefficient are generally difficult to address in a physically meaningful way. Future improvements in orbital aerodynamic science may enable to establish a better correlation of this latter with the environment, the spacecraft thermal model, and the material selection.
Once the main body and the control surfaces have been identified, the algorithm computes the aerodynamic coefficients for a range of possible satellite configurations by rotating each panel about its longitudinal axis. Since no particular restriction is imposed on the panel movement, the number of possible permutations can be high, especially if small angular steps ($\vartheta_{s}$) are considered. In order to reduce the computational effort, only a limited number of possible configurations is selected by the PCA. For each panel, a vector $ap_{i}$ containing three angular positions is defined according to the following definition:
\begin{equation}\label{eq: ap_i}
ap_{i}= \left[\lfloor\vartheta_{p,i_{k}}/\vartheta_{s}\rfloor\vartheta_{s}-\vartheta_{s},
\lfloor\vartheta_{p,i_{k}}/\vartheta_{s}\rfloor\vartheta_{s},
\lfloor\vartheta_{p,i_{k}}/\vartheta_{s}\rfloor\vartheta_{s}+\vartheta_{s}\right]^{T}  \quad\text{for}\quad 
i = 1,...,4
\end{equation}
where $\vartheta_{p,i_{k}}$ is the estimated angle of deflection of the i-th panel at the time step $t_{k}$, $\vartheta_{s}$ is the reference rotation angle of the panels used to perform the computation of the aerodynamic coefficients, and  $\lfloor x\rfloor $ 
indicates the floor function for $x = \vartheta_{p,i_{k}}/\vartheta_{s} $. The selection of the angles of deflection for each panel is subjected to the definition of a saturation limit for the actuators ($\pm\vartheta_{max}$), so that:
\begin{equation}\label{eq: ap_i with limits}
ap_{i} = \left\{ap_{i1}, ap_{i2}, ap_{i3}:  -\vartheta_{max}\le ap_{i1}, ap_{i2}, ap_{i3}\le \vartheta_{max}\right\}
\end{equation}
A rectangular grid ($P_{R}$) containing the combinations of the angular positions specified in $ap_{1}$, $ap_{2}$, $ap_{3}$ and $ap_{4}$ is then created, so that:
\begin{equation}\label{eq: PR matrix}
P_{R} =
\begin{bmatrix}
ap_{11}&ap_{21}&ap_{31}&ap_{41}\\
ap_{12}&ap_{21}&ap_{31}&ap_{41}\\
\vdots&\vdots&\vdots&\vdots\\
ap_{12}&ap_{23}&ap_{33}&ap_{43}\\
ap_{13}&ap_{23}&ap_{33}&ap_{43}\\
\end{bmatrix}
\end{equation}
Each row in the $P_{R}$ matrix defines a configuration of the aerodynamic panels. For any of these, the PCA reads the angles of deflection associated with the panels and rotates the coordinates of the corresponding elements in the mesh by applying the rotation matrices described in Eq. \ref{dcm: panels rotation}.
The PCA then proceeds to the computation of the dimensional aerodynamic coefficients in Eq. \ref{aero_torque}. For the n-th configuration considered, these are given by:
\begin{equation}\label{dimensional aero coeff}
\boldsymbol{C_{M_{d}}}^{(n)} = \boldsymbol{C_{M}}^{(n)}S_{ref}\ell_{ref}= \sum^{M_{tot}}_{j = 1}
\left(\boldsymbol{r_{j}} -\boldsymbol{r_{CoM}} \right)\times\left(c_{\tau j}^{(n)}\boldsymbol{\hat \tau_{j}}^{(n)}-c_{p j}^{(n)}\boldsymbol{\hat n_{j}}^{(n)}\right)S_{j}
\end{equation}
where $\boldsymbol{r_{j}}$ is the vector that describes the distance between the barycenter of the j-th element in the mesh and the geometric centre of the satellite, $\boldsymbol{r_{CoM}}$ is the position vector of the satellite CoM measured from the satellite geometric centre, $c_{\tau j}^{(n)}$ and $c_{p j}^{(n)}$ are respectively the aerodynamic shear stress and normal pressure coefficients, $\boldsymbol{\hat n_{j}}^{(n)}$ is the outward unit normal vector, $S_{j}$ is the area  and  $\boldsymbol{\hat \tau_{j}}^{(n)}$ is the unit tangent vector of the j-th element considered in the mesh.

The definition of the aerodynamic coefficients $c_{pj}^{(n)}$ and $c_{\tau j}^{(n)}$ is not necessarily unique. As discussed in \cref{engineering variables} different analytical expressions are given according to the assumed re-emission mechanism of the particles and the GSI model employed to describe it. For the application discussed in this paper, these will be chosen considering the limited performance achievable by common materials. In these regards, Sentman's model \citep{Sentman1961} is widely used to describe diffuse particle re-emissions from contaminated surfaces and will therefore be adopted: 
\begin{subequations}
\begin{gather}\label{eq: cp_sentman}
\begin{split}
c_{pj}^{(n)} =\; & \frac{\cos(\delta_{j}^{(n)})}{\sqrt{\pi}s}e^{-s^{2}\cos^{2}(\delta_{j}^{(n)})} +
\left(\frac{1}{2s^{2}}+\cos^{2}(\delta_{j}^{(n)})\right)\left[1+\erf(s\cos(\delta_{j}^{(n)}))\right]+\\
&+\frac{1}{2}\sqrt{\frac{2}{3}\left[1 + \alpha_{T,j}\left(\frac{T_{w}}{T_{i}} - 1\right) \right]} \left\{\sqrt{\pi}\cos(\delta_{j}^{(n)}) \left[1 + \erf(s\cos(\delta_{j}^{(n)}))\right]+ \right.\\ & \left.+ \frac{1}{s}e^{-s^{2}\cos^{2}(\delta_{j}^{(n)})}\right\}
\end{split}\\
\label{eq: ctau_sentman}
c_{\tau j}^{(n)} = \frac{\sin(\delta_{j}^{(n)})}{\sqrt{\pi}s}\left\{e^{-s^{2}\cos^{2}(\delta_{j}^{(n)})} +\sqrt{\pi}s\cos(\delta_{j}^{(n)})\left[1+\erf(s\cos\delta_{j}^{(n)})\right]\right\}
\end{gather}
\end{subequations}
where $s$ is the molecular speed ratio (see Eq. \ref{molecular speed ratio}), $T_{w}$ is the wall temperature, $\alpha_{T,j}$ is the thermal accommodation coefficient defined in Eq. \ref{alpha_accom}, $T_{i} = \frac{2}{3}s^{2}\,T_{alt}$ is the kinetic temperature of the incident flow expressed as a function of the thermospheric temperature at the specified altitude ($T_{alt}$),
and $\delta_{j}^{(n)}$
 is the angle between the incident flow and the normal to each surface element in the mesh for the n-th configuration considered. Approximated values of $s$ and $T_{alt}$ are obtained by linear interpolation of some averaged reference values provided by the NRLMSISE-00 model  \citep{Picone2002} for the space weather conditions at the epoch selected (see Tab. \ref{table: environmental parameters}).
\begin{table}[hbt!]
\caption{\label{tab:table1} Averaged reference values of $\boldsymbol{s}$ and $\boldsymbol{T_{alt}}$.}
\centering
\begin{tabular}{ccc}
\hline
\rule{0pt}{2ex}
\textbf{Altitude [km]}& \textbf{Molecular speed ratio} & \textbf{Temperature at altitude  [K]}\\\hline
\rule{0pt}{3ex}
200 & 10.6657 & 669.3270\\
210 & 10.4548 & 676.1180\\
220 & 10.2676 & 681.0724\\
230 & 10.1002 & 684.7046\\
240 & 9.9499 &687.3797\\
250 &9.8145&689.3585\\
260 &9.6921&690.8283\\
270 &9.5806&691.9243\\
280 &9.4781&692.7448\\
290 &9.3826&693.3612\\
\hline
\end{tabular}\label{table: environmental parameters}
\end{table}
To reduce the number of computations to be performed, the dimensional aerodynamic coefficients in Eq. \ref{dimensional aero coeff} are computed separately for the mesh elements belonging to the main body ($\boldsymbol{C_{M_{d,B}}}$) and the mesh elements belonging to the control panels rotated at the angles specified in $P_{R}$ ($\boldsymbol{C_{M_{d,P}}}^{(n)}$). The contribution due to the satellite main body is determined by the attitude of the satellite with regards to the flow and it is therefore the same for all the configurations considered. For the n-th configuration, the expected aerodynamic control torque is then computed as:
\begin{equation}
\boldsymbol{T_{a}}^{(n)} = q_{d}\left(\boldsymbol{C_{M_{d,B}}} + \boldsymbol{C_{M_{d,P}}}^{(n)}\right)
\end{equation}
The angular position of the fins selected by the PCA is the one for which the induced aerodynamic torques best match the commanded control torques while minimising as much as possible the induced aerodynamic drag. The desired configuration is the one that produces the minimum Euclidean distance ($d_{L_{2}}$) between the vectors $\boldsymbol{T_{ref}}$ and $\boldsymbol{T_{exp}}$:
\begin{equation}\label{min_eucl_dist}
\begin{split}
\left[\vartheta_{p,1}^{(n)}, \vartheta_{p,2}^{(n)}, \vartheta_{p,3}^{(n)}, \vartheta_{p,4}^{(n)}\right] \Leftrightarrow d_{L_{2}}^{(n)}\left(T_{ref},T_{exp}^{(n)}\right) = &\: \sqrt {\sum _{ii=1}^{l}  \left( T_{ref,ii}-T_{exp,ii}^{(n)}\right)^2 }\equiv \min\left[ d_{L_{2}}\left(T_{ref},T_{exp}\right) \right]
\end{split}
\end{equation}
with $\boldsymbol{T_{ref}} = [\boldsymbol{u_{a,k}}, P_{N}]$ and $\boldsymbol{T_{exp}}^{(n)} = [\boldsymbol{T_{a}}^{(n)}, \gamma_{p}| P_{R_{n,*}}|]$, where $P_{N}$ is the vector of the panels angles in the nominal low drag configuration expressed in radians, $| P_{R_{n,*}}|$ is the absolute value in radians of the angles of deflection of the four panels as they are defined for each configuration in $P_{R}$ and $\gamma_{p}$ is a weighting coefficient. If multiple configurations providing the same torque are identified, the one that requires the least movement of the control surfaces is selected in order to reduce the disturbance introduced in the total system dynamics.
\section{Validation of the PCA algorithm and discussion}\label{Validation of the PCA algorithm and discussion}
\begin{figure}\begin{centering}
\includegraphics[width=1\textwidth]{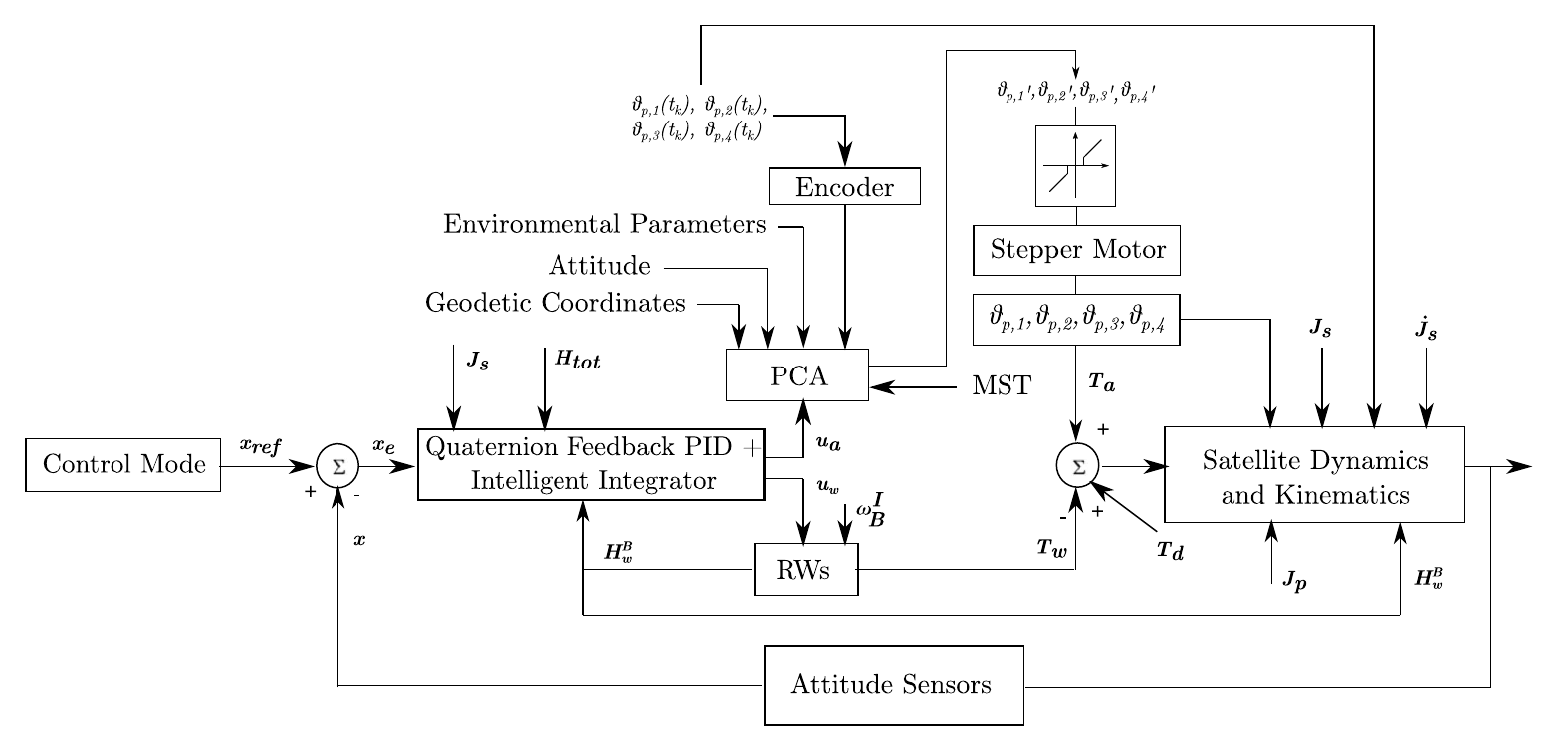}
\caption{Control scheme for the PCA validation and the Monte Carlo analysis.}
\label{combined}
\end{centering}
\end{figure}
This section discusses the capability of the PCA algorithm described in \cref{AFAAC} to select, at each time step $t_{k}$, a configuration of the panels providing the required torque $\boldsymbol{u_{a}(t_{k})}$ received in input. For this purpose, the aerodynamic control torque induced about the desired control axes by the selected angles of deflection of the actuators is compared with the command torque computed by the discrete time implementation of a quaternion feedback PID controller with an intelligent integrator \citep{Wie1989,Bang2003} for a sampling time $t_{s}= 1\text s$. Results refer to the duration of a representative combined pointing manoeuvre for which aerodynamic torques are used to control the roll dynamics, and RWs in a tetrahedron configuration are used to stabilise the pitch and yaw dynamics (Fig. \ref{combined}). An initial 250 km circular orbit inclined at 51.6$\degree$ is assumed and the initial mean anomaly is set to $\nu_{M} = 0\degree$. The
selected reference epoch for the simulations is 2020 April 15 04:50:00, for which low solar activity is expected in solar cycle 25. The associated values of the solar and magnetic proxies are selected according to the standards specified by ISO 1422:2013  \citep{ISO2013}. Attitude propagation is performed keeping into account the contribution due to the major environmental disturbances experienced by satellites in VLEO: solar radiation pressure torques are modelled assuming a specular reflectivity coefficient of $c_{s} = 0.15$ and a diffuse reflectivity coefficient equal to $c_{d} = 0.25$; the residual dipole moment is assumed equal to $m_{b} = 0.01\text{ Am\textsuperscript{2}}$ and the Earth's magnetic field is modelled according to the International Geomagnetic Reference Field (IGRF-12). Orbital perturbations include aerodynamic and solar radiation pressure accelerations and zonal gravity harmonics up to $J_{4}$. Simulations are performed by means of 6-DOF attitude and orbit propagator that, unlike the controller, is implemented in continuous rather than in discrete time.
Validation is performed including as many sources of uncertainty discussed in \cref{orbit_var} and \cref{engineering variables} as possible:
\begin{enumerate}
\item{The PCA algorithm is designed to use approximated and interpolated averaged values of the environmental parameters.
The actual aerodynamic control torque induced by the panels is determined by using the NRLMSISE-00 atmospheric
model  \citep{Picone2002} and by performing precise estimation of local solar time.}
\item{Thermospheric winds are neglected in the decision process of the PCA (\cref{Relative velocity estimation}), but these are included in the determination of the angle of attack and sideslip when determining the actual aerodynamic torque provided by the panels. The HWM93 model  \citep{Hedin1996} is used for this purpose.}
\item{Differently from the PCA, the simulation environment assumes partial particle accommodation. Due to the low altitude orbit considered and the performance of typical materials a realistic value of $\alpha_{T} = 0.95$ is employed.}
\item{The four panels are constrained to rotate with an angular speed of $\omega_{p,i} = 0.8\degree/\text{s}$ and they are subject to a saturation limit of $\vartheta_{max} = \pm 60\degree$. The angular step used to perform the computation in the PCA is $\vartheta_{s}=\pm4\degree$. A minimum angular step for the plant ($\vartheta_{pl}= \vartheta_{s}$) is also introduced, so that if the commanded $\vartheta_{p,i}<\vartheta_{pl}$, the i-th panel would be kept in the nominal configuration. }
\item{The performance of off-the-shelf components available for platforms similar to the one considered is reproduced. The approach followed is based on GomSpace's characterisation of the UKF attitude determination capability of SOAR. Attitude knowledge error is modeled as gaussian noise corresponding to the filtered 2$\sigma$ value obtainable with the Epson M-G370 gyro suite and a three-axis magnetometer at 200 km during periods of eclipse ($0.42\degree$) and Sun ($0.43\degree$). Analogously, errors in the determination of the panels angular positions are modelled as gaussian noise according to the performance of the AM4096 - 12 bit angular magnetic encoder, by using as a conservative estimation position errors of $\pm0.2\degree$. Inertial angular body rates are recovered from gyro measurements as in Ref. \cite[pp. 266-270]{Wertz1999}: white gyro noise is modelled as a function of the angle random walk expected for the M-G370 gyro suite. A $\pm10\%$ uncertainty is introduced on the angular steps of the panels, according to the performance expected for a Faulhaber two-phase stepper motor.}
\item{The impact of the rotating appendages on the satellite dynamics is included in the simulated system behaviour.}
\end{enumerate}
The main geometric characteristics of SOAR are summarised in Tab. \ref{table: geometric characteristics}, where the coordinates of the CoM are measured from the satellite geometric centre according to the body axes convention. Due to its convex geometry and the extension of the control surfaces with regards to the main body dimensions, panel shadowing can be reasonably neglected. 
\begin{table}[hbt!]
\caption{ SOAR's geometric features. Symbols $\boldsymbol{\ell,w\text{ and }t}$ respectively indicate length, width and thickness.}
\centering
\begin{tabular}{ccccc}
\hline
\rule{0pt}{2ex}
\textbf{Main bus [m]} & \textbf{Panels [m]} & \textbf{Panel support [m]} & \textbf{CoM location [m]}& \textbf{Inertia tensor [kg m\textsuperscript{2}]} \\\hline
\rule{0pt}{3ex}
$\ell = 0.3660$&$\ell = 0.3660$&$\ell = 0.007$&\multirow{3}{*}{$\scriptstyle\boldsymbol{r_{CoM}} =\begin{bmatrix}\scriptstyle{-22.01e^{-3}}\\\scriptstyle{-0.999e^{-3}}\\\scriptstyle{0.77e^{-3}}\end{bmatrix}$}&\multirow{3}{*}{$\scriptstyle\boldsymbol{J_{s}} = \setlength\arraycolsep{1pt} \begin{bmatrix}\scriptstyle{0.0543}&\scriptstyle{1.00e^{-5}}&\scriptstyle{-7.79e^{-6}}\\\scriptstyle{1.00e^{-5}}&\scriptstyle{0.0627}&\scriptstyle{-2.21e^{-6}}\\\scriptstyle{-7.79e^{-6}}&\scriptstyle{-2.21e^{-6}}&\scriptstyle{0.0627}\end{bmatrix}$}\\
$w=0.1$&$w=0.060$&-\\
-&$t=0.001$&-\\
\hline
\end{tabular}\label{table: geometric characteristics}
\end{table}
\begin{figure}\begin{centering}
\includegraphics[width=1\textwidth]{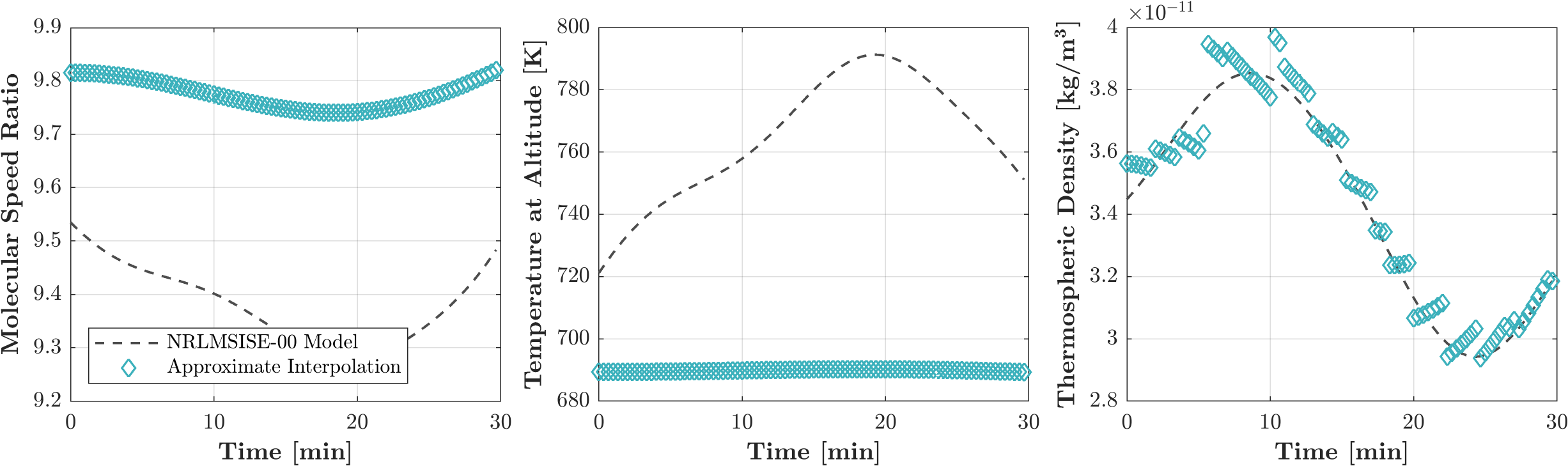}
\caption{Values of the environmental parameters as they are computed by: 1) the simple linear interpolation  performed by the PCA and 2) the NRLMSISE-00 atmospheric model  \citep{Picone2002}.}
\label{EnvironmentalParameters_Uncertainties}
\end{centering}
\end{figure}

\begin{figure}[hbt!]\begin{centering}
\includegraphics[width=1\textwidth]{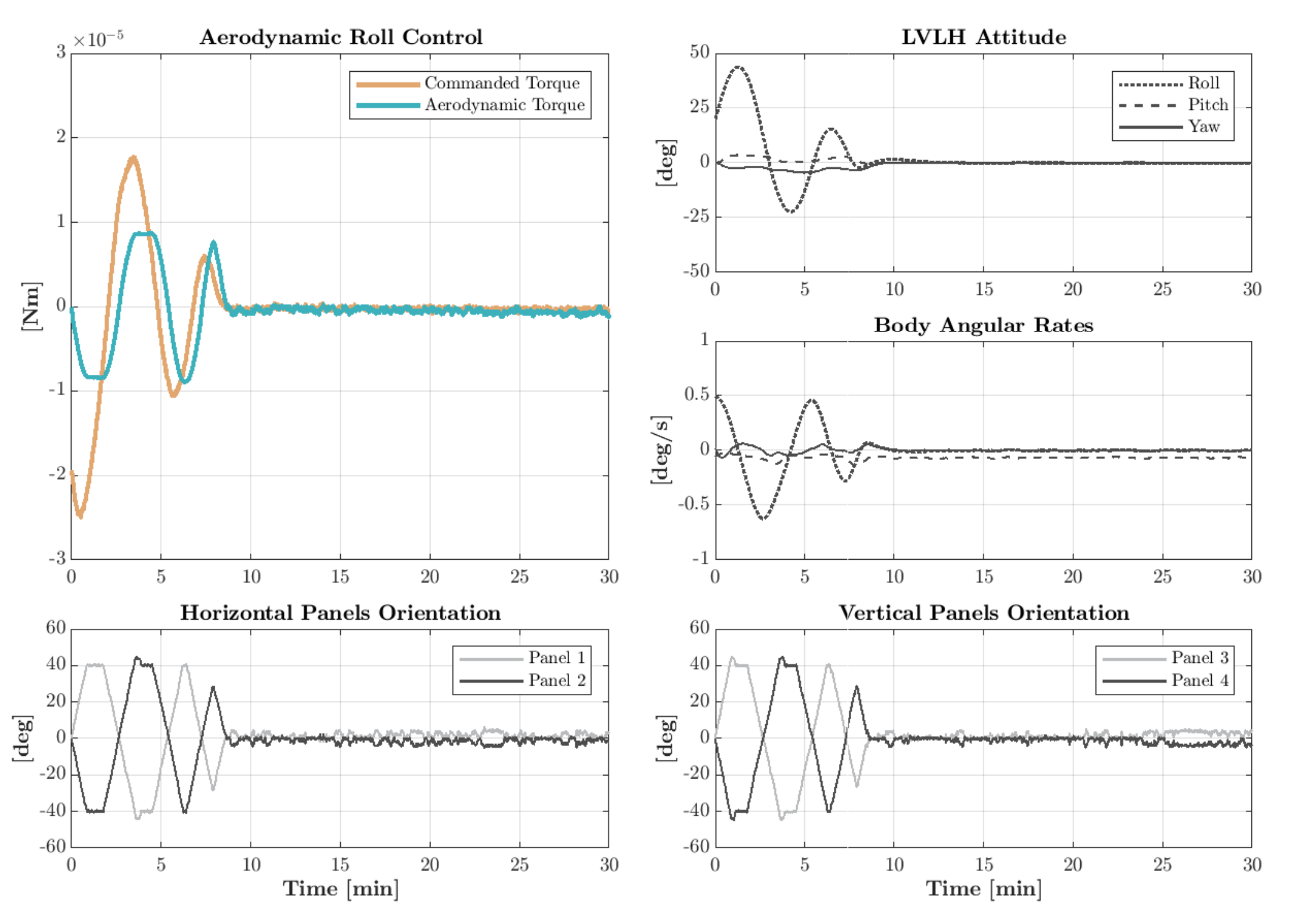}
\caption{Validation of the PCA algorithm for the case of a roll aerodynamic attitude control manoeuvre.}
\label{Roll_PCA_Validation}
\end{centering}
\end{figure}

 Fig. \ref{EnvironmentalParameters_Uncertainties} shows the values of the environmental parameters computed using the approximate decision process of the PCA algorithm and the more accurate NRLMSISE-00 atmospheric model  \citep{Picone2002} over the duration of the pointing manoeuvre shown in Fig. \ref{Roll_PCA_Validation}. Any discrepancy between the values expected by the PCA algorithm and the actual environmental conditions represents a source of uncertainty against which the algorithm must show robustness. As expected, the estimation of the molecular speed ratio and the temperature at the current altitude performed by the algorithm is fairly constant. 
The quick and simplified logic designed is not capable of capturing short-scale temporal and spatial variations, accounted for the NRLMSISE-00 model. 
As evidenced by the right plot in Fig. \ref{EnvironmentalParameters_Uncertainties}, discrepancies in the estimation of $\rho$ may cause the algorithm to partially underestimate or overestimate the control authority effectively achievable. Differences with the values predicted by the atmospheric model are due to the fact that a very limited range of reference latitudes and local solar time values are considered for the interpolation of the approximate density.
\begin{figure}\begin{centering}
\includegraphics[width=0.9\textwidth]{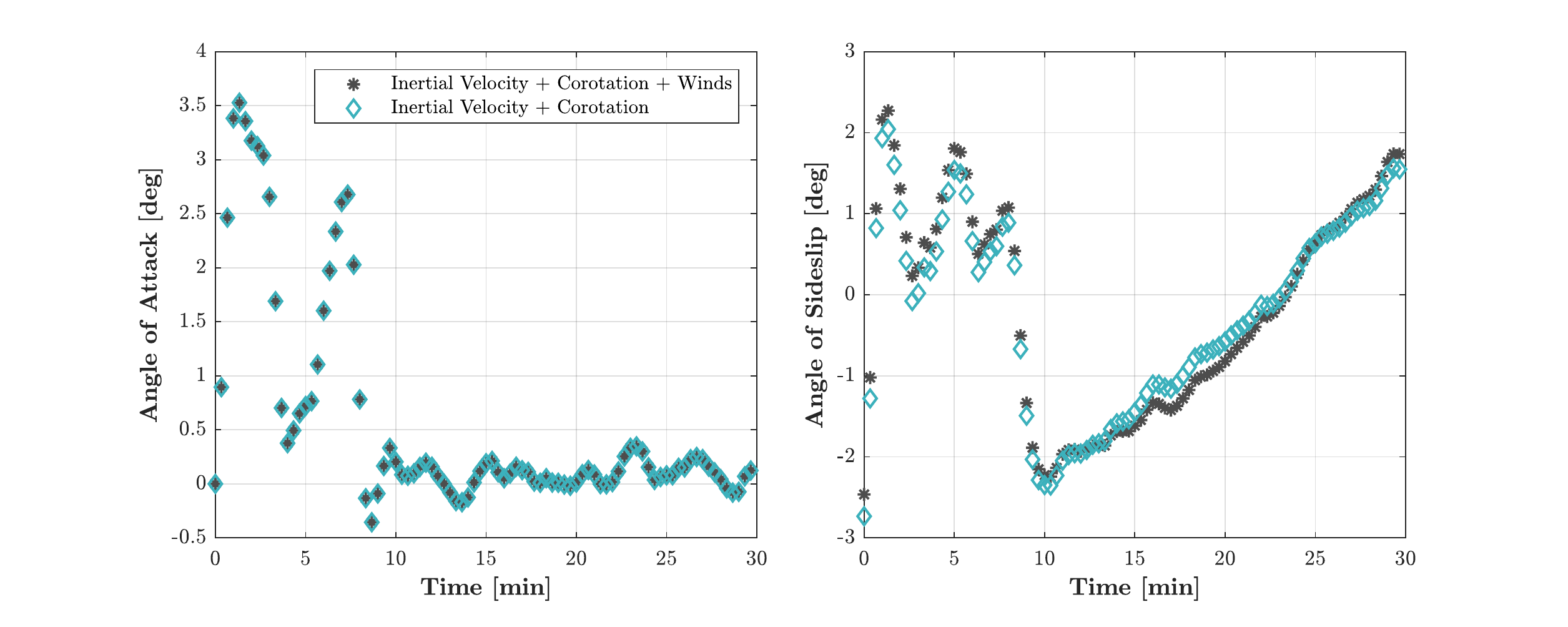}
\caption{Attitude estimation considering: 1) the inertial velocity and the atmospheric corotation vectors (PCA algorithm); 2) the inertial velocity, the atmospheric corotation, and the horizontal winds vector (induced aerodynamic output).}
\label{Attitude_Uncertainties}
\end{centering}
\end{figure}
 Moreover, Fig. \ref{Attitude_Uncertainties} shows that when horizontal winds are included in the computation of the aerodynamic torque induced by the panels, discrepancies with the PCA algorithm are observed in the computation of the angle of sideslip.
Inclusions of vertical winds, which are here neglected, are expected to introduce some small uncertainties also in the determination of the angle of attack. 

Given these premises, Fig. \ref{Roll_PCA_Validation} shows the behaviour of the PCA algorithm while performing an aerodynamic attitude control manoeuvre to stabilise the motion about the roll axis from an initial offset of $\phi=20\degree$, and an initial body angular rate of $0.5\degree/\text{s}$. Aerodynamic control authority is maximised by limiting the decision process of the PCA to panels counter rotated configurations. The control signal results from setting alignment with the Local Vertical Local Horizontal reference frame \cite[pp. 36--37]{Markley2014} as the desired final state.
 The aerodynamic control torque induced by the configuration of the panels selected by the PCA at each time step (blue solid line in Fig. \ref{Roll_PCA_Validation}, left) is compared against the corresponding command torque computed by the quaternion feedback PID controller (orange solid line in Fig. \ref{Roll_PCA_Validation}, left). 
Despite the uncertainties described above, the trend observed in the command torque is qualitatively reproduced by the aerodynamic torque induced by the panels. The quantitative differences observable at the beginning of the manoeuvre are likely to derive from the requirement imposed on the speed of rotation of the panels.

\section{Robustness analysis}\label{MCanalysis}
\begin{figure}[hbt!]\begin{centering}
\includegraphics[width=1\textwidth]{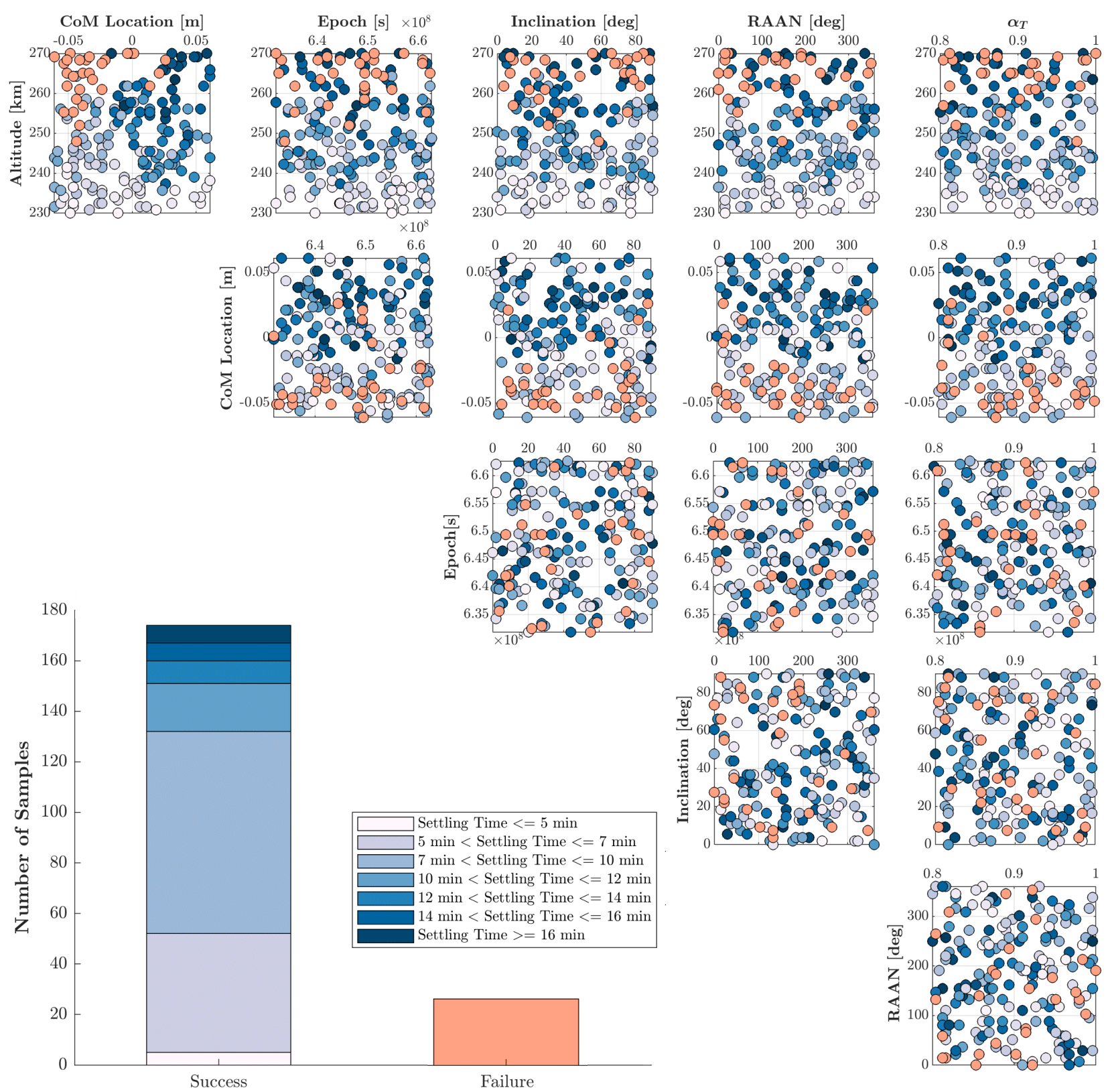}
\caption{Top right: covariance plots of the independent variables values selected by the random process of the Monte Carlo simulation. Bottom left: bar plots of the failed (orange) and successful (blue scale) samples considered.}
\label{fig:scatterplots}
\end{centering}
\end{figure}

\begin{figure}\begin{centering}
\includegraphics[width=1\textwidth]{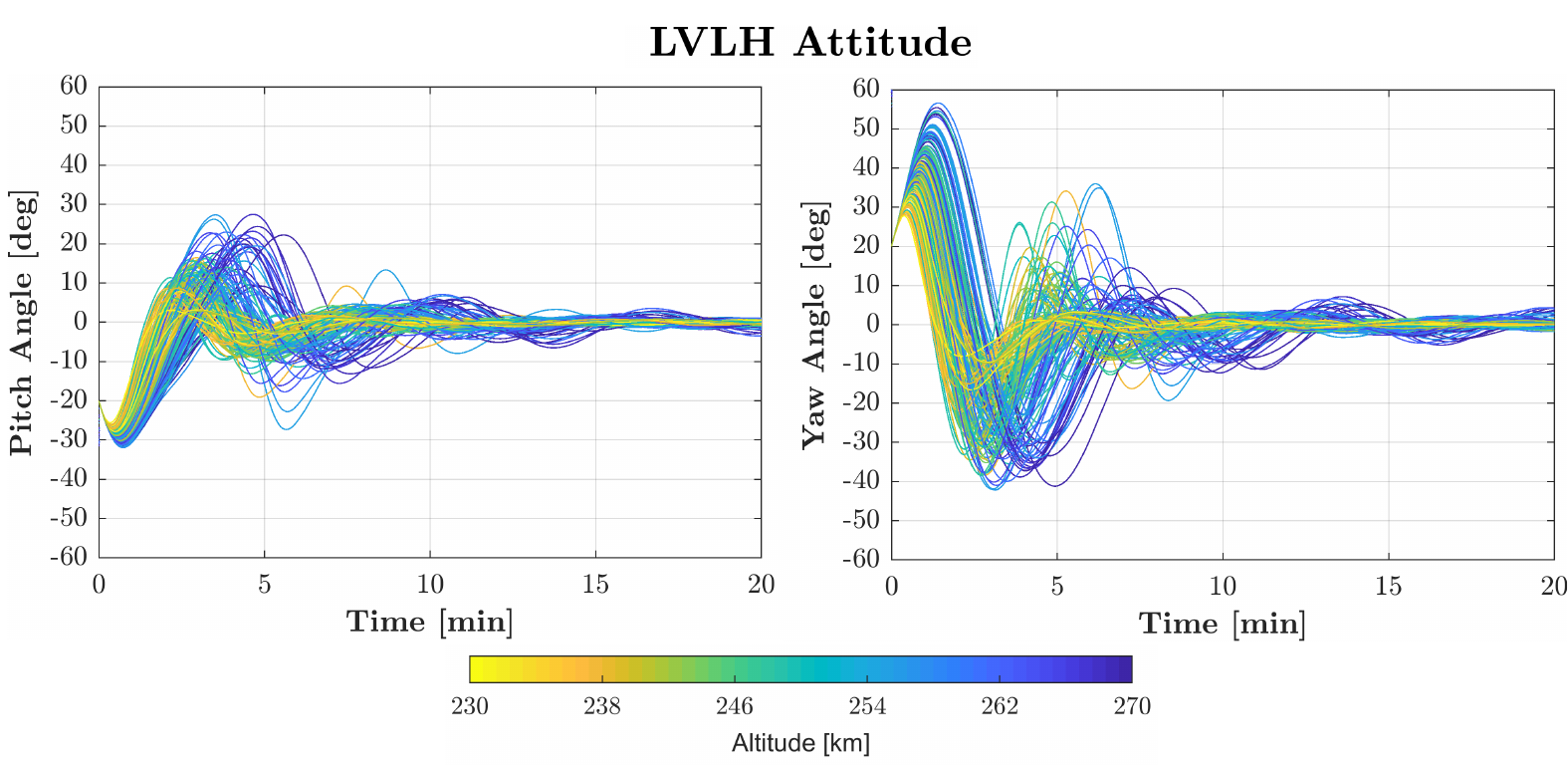}
\caption{Time histories of the pitch and yaw angles for each sample cases considered in the Monte Carlo analysis. Color coding is attributed according to the Monte Carlo simulation altitudes.}
\label{fig:pitchyawdynamics}
\end{centering}
\end{figure}

A Monte Carlo simulation was performed to assess the sensitivity of the algorithm in the presence of uncertain modelling and to identify under which circumstances the simple aerodynamic controller proposed may fail. 
The parameters utilised in the simulation are shown in Tab. \ref{table: MCrange}.
The implementation of the PCA algorithm, the system considered, and the lookup table used to predict density are unvaried and coincide with the ones discussed in the previous sections. A combined manoeuvre for which aerodynamic panels are commanded to control the pitch and yaw dynamics and RWs are used to stabilise the roll motion, is assumed (Fig. \ref{combined}). For the aerodynamic coefficient computation, the PCA assumes that the CoM is located at its nominal position (see Tab. \ref{table: MCrange}) for all the case studies considered: the effect of ignoring the CoM "true" position can thus be accordingly addressed in the results discussion. The quaternion feedback PID \citep{Bang2003} gains are determined by using an optimal linear quadratic regulator approach \citep{Heidecker2009}. For all the cases considered these are derived by always selecting the same weighting coefficients for the state space ($w_{Q} = [1,1,1]^{T}$) and the control input ($w_{R} = [1e^{11},1e^{12},1e^{12}]^{T}$) matrices, assumed diagonal. The range of values for each variable (see Tab. \ref{table: MCrange}) has been defined with the specific purpose of probing the failure conditions for the controller. For practical applications orbital parameters such as RAAN and orbit inclination are typically known with considerably higher confidence. Similarly, PID gains can be tuned for altitude uncertainties that are smaller than those considered. CoM knowledge accuracy is generally in the order of millimeters/centimeters. A wider range of altitudes, however, offers the chance to assess to what extent simple control laws, such as a quaternion feedback PID, can be considered robust when used in combination with the proposed algorithm to implement aerodynamic  control. The question appears relevant, since PIDs are in practice the industrial standard for most space applications. Analogously, reduced CoM-CoP distances may provide information on some design criteria that may be relevant especially during quiet solar cycle conditions. Since off-nominal conditions are expected to have an impact especially on the capability of damping the platform residual rates, a challenging scenario was considered. Aerodynamic control was initiated with an initial offset of $\varphi = -20\degree$  in pitch and $\psi = 20\degree$ in yaw and angular body rates of $-0.5\degree/\text{s}$ and $0.5\degree/\text{s}$  about the pitch and yaw axes, respectively. 
The target state imposed is alignment with the LVLH reference frame \cite[pp. 36--37]{Markley2014}.
For each instance considered, the manoeuvre was considered failed if saturation of at least one wheel in the assembly occured before the satellite attitude was stabilised. Vice versa, the manoeuvre was assumed to be successfully achieved if the satellite attitude was coarsely stabilised with steady state errors below $3\degree$ for more than 30 s without incurring in RWs saturation. In the presence of larger steady state errors but no RWs saturation, the manoeuvre was considered achieved but with degraded performance.
\begin{table}[hbt!]
\caption{Ranges of variation of the independent variables used to perform the Monte Carlo analysis.}
\centering
\begin{tabular}{lcc}
\hline
\rule{0pt}{3ex}
\textbf{Variable} &\textbf{Nominal Value} & \textbf{Range} \\\hline
\rule{0pt}{3ex}
Epoch & 2020 Apr 15 04:50:00 &[2020 Jan 1, 2021 Jan 1]\\
\rule{0pt}{2ex} $\alpha_{T}$&$0.95$&$[0.8,1]$\\
\rule{0pt}{2ex}   RAAN [deg]&0&$[0,360]$\\
\rule{0pt}{2ex}   Inclination [deg]&51.6&$[0,90]$\\
\rule{0pt}{2ex}   Altitude [km]&250&$[230,270]$\\
\rule{0pt}{2ex}   CoM location [m]&$-0.0220086$&$[-0.061,0.061]$\\
\hline
\end{tabular}\label{table: MCrange}
\end{table}
 Results obtained by random selection of the independent variables for 200 samples are shown in Fig. \ref{fig:scatterplots} and Fig. \ref{fig:pitchyawdynamics}. In Fig. \ref{fig:scatterplots}, orange dots indicate failed samples, whilst dots with a color gradient ranging from white to dark blue identify successful samples. Darker colors are associated to successful samples with longer settling times and thus inferior performance. The outcome of the simulation is displayed in such a manner that covariance of the independent variables can be discussed. As expected, the majority of failures occur when the aerodynamic control authority experienced is substantially lower than that predicted by the controller. 
This condition occurs when two conditions are verified: 1) operations are conducted at increased altitudes, and 2) a platform with a reduced offset between the CoM and the CoP is utilised (i.e. the CoM is located behind the geometric centre of the bus). The inability to damp the residual body rates does not only depend on the reduced aerodynamic control authority on itself, but also by the fact that the utilised set of gains were selected for more favourable control conditions (nominal parameters value in Tab. \ref{table: MCrange}). When the aerodynamic authority reduces, if the gains are too small, energy damping is not fast enough and saturation of the reaction wheels quickly occurs. This suggests a conclusion that is somewhat expected: to achieve comparable performance in different conditions using a simple PID, gains scheduling is required. However, the selection of more favourable conditions for at least one parameter between altitude and the CoM location seems to be enough to assure the robustness of the algorithm under the uncertainties considered. Lower altitudes and/or longer CoM-CoP distances result in increased aerodynamic torques and thus better performance. In this regard, it is interesting to notice how moving the CoM forward of the geometric centre of the main bus may efficiently compensate for density reduction at higher altitudes. The relevant uncertainties affecting the algorithm decision process translates to longer settling times but does not result in the failure of the manoeuvre. For the CubeSat satellite class, however, any solution for shifting forward the CoM or to move back the CoP can be only utilised after orbit insertion: launch requirements impose the CoM to be located within 2 cm from the geometric centre. Since the diurnal bulge is the most relevant feature for density variations on a horizontal plane, the effect of coupling with seasonal uncertainties is generally small, as evidenced by the altitude and CoM plots against epoch. Orbit inclination, RAAN and $\alpha_{T}$ overall seem to have a smaller impact on the sensitivity of the controller. Results, however, are expected to vary according to solar activity conditions considered. For intense solar irradiance, the behaviour of the algorithm at higher altitudes and reduced CoM-CoP distances is expected to improve due to a general increase of atmospheric density over the VLEO range (see Fig. \ref{with_altitude}). On the other hand, inclination is expected to have a more substantial role especially at higher latitudes, where thermospheric winds may reach velocities of 400 m/s \citep{Doornbos2011}. In general, even better results are expected to be observed for less demanding control manoeuvres.
Overall results seem to suggest that during quiet solar activity is recommended to perform operations below 250 km, where atmospheric density is high enough for manoeuvring. The capability to attain the target attitude for 87\% of the samples, despite the large width of variation of the independent variables considered, attests the robustness of the controller.

 Some of the considerations that have been made here are confirmed by the time histories of the pitch and yaw dynamics displayed in Fig. \ref{fig:pitchyawdynamics}. In this case, color codes are used to discuss the impact of the altitude on accuracy, overshoot, and settling time of the manoeuvre. As expected, generally worse performance is associated with manoeuvring at higher altitudes. Reduced overshoot, smaller settling times, and improved accuracy are instead observed especially for samples at 230-240 km. Some deviations from this general trend are however observed, as a result of the selection of the other independent variables in the problem. To an extent, it is possible that higher steady state errors observed in some cases may result from utilising a set of gains that was not optimal for the orbital conditions. In the worst case displayed, however, the satellite attitude is constrained between $\pm5\degree$ from the desired final target. Under the current technological level (available performance of materials) and low solar activity conditions, only coarse aerodynamic pointing can be expected. However, in the near future potential improvements may be achieved with the utilisation of quasi-specularly reflecting materials, improved hardware and software specifications, and with an increased confidence in estimating density and thermospheric winds.
\section{Conclusions}\label{Conclusion and final remarks}
The algorithm proposed to actuate four independent control surfaces to achieve active aerodynamic control was validated by assuming representative case studies for combined aerodynamic and reaction wheels attitude pointing. The results of the  Monte Carlo analysis performed show that, even when a simple quaternion feedback PID is used to determine the input control signal, coarse pointing is achieved and the algorithm ensures robustness against uncertainties, inaccurate environmental modelling, and attitude hardware limitations. Critical conditions for the utilisation of aerodynamic control during quiet solar activity are presented when the control authority significantly reduces (high VLEO altitudes with small CoM-CoP offsets) and, at the same time, the set of gains selected for the quaternion feedback PID control law is not appropriate to assure fast damping. Overall, operations in VLEO may significantly benefit from an advantageous use of the aerodynamic environment to perform attitude control tasks. 

\section*{Acknowledgements}
The authors of this paper are grateful to the members of the DISCOVERER project and to Dr David Mostaza Prieto for their suggestions, and for sharing their research experience with them.
This project has received funding from the European
Union's Horizon 2020 research and innovation programme
under grant agreement No 737183. This publication reflects
only the view of the authors. The European Commission
is not responsible for any use that may be made of the
information it contains.

\bibliography{library.bib}

\end{document}